\documentclass[12pt,preprint]{aastex}

\newcommand{\kms}{\ifmmode {\rm km\ s}^{-1} \else km s$^{-1}$\fi}
\newcommand{\Msun}{\ifmmode {\rm M}_{\odot} \else M$_{\odot}$\fi}
\newcommand{\Lsun}{\ifmmode {\rm L}_{\odot} \else L$_{\odot}$\fi}
\newcommand{\qo}{\ifmmode q_{\rm o} \else $q_{\rm o}$\fi}
\newcommand{\Ho}{\ifmmode H_{\rm o} \else $H_{\rm o}$\fi}
\newcommand{\ho}{\ifmmode h_{\rm o} \else $h_{\rm o}$\fi}

\newcommand{\gtsim}{\raisebox{-.5ex}{$\;\stackrel{>}{\sim}\;$}}
\newcommand{\vFWHM}{\ifmmode v_{\mbox{\tiny FWHM}} \else
                    $v_{\mbox{\tiny FWHM}}$\fi}
\newcommand{\CCF}{\ifmmode F_{\it CCF} \else $F_{\it CCF}$\fi}
\newcommand{\ACF}{\ifmmode F_{\it ACF} \else $F_{\it ACF}$\fi}
\newcommand{\Halpha}{\ifmmode {\rm H}\alpha \else H$\alpha$\fi}
\newcommand{\Hbeta}{\ifmmode {\rm H}\beta \else H$\beta$\fi}
\newcommand{\Hgamma}{\ifmmode {\rm H}\gamma \else H$\gamma$\fi}
\newcommand{\Hdelta}{\ifmmode {\rm H}\delta \else H$\delta$\fi}
\newcommand{\Lya}{\ifmmode {\rm Ly}\alpha \else Ly$\alpha$\fi}
\newcommand{\Lyb}{\ifmmode {\rm Ly}\beta \else Ly$\beta$\fi}
\newcommand{\HeI}{\ifmmode {\rm He}\,{\sc i}\,\lambda5876 \else 
	          He\,{\sc i}\,$\lambda5876$\fi}
\newcommand{\HeII}{\ifmmode {\rm He}\,{\sc ii}\,\lambda4686 \else 
	           He\,{\sc ii}\,$\lambda4686$\fi}

\newcommand{\hei}{He\,{\sc i}}
\newcommand{\heii}{He\,{\sc ii}}

\newcommand{\feii}{Fe\,{\sc ii}}
\newcommand{\feiii}{Fe\,{\sc iii}}

\newcommand{\ciii}{\ifmmode {\rm C}\,{\sc iii} \else C\,{\sc iii}\fi}
\newcommand{\civ}{C\,{\sc iv}}

\newcommand{\nii}{N\,{\sc ii}}

\newcommand{\oiii}{O\,{\sc iii}}
\newcommand{\ob}{[O\,{\sc iii}]\,$\lambda \lambda 4959,5007$}

\newcommand{\sigbl}{$\sigma_{\rm blue}$}
\newcommand{\Flamunit}{erg s$^{-1}$\,cm$^{-2}$\,\AA$^{-1}$}

\shorttitle{Systematics in Single Epoch Black Hole Mass Measurements}
\shortauthors{}

\begin{document}

\title{Systematic Uncertainties in Black Hole Masses Determined from Single Epoch Spectra}

\author{ Kelly~D.~Denney\altaffilmark{1},
         Bradley~M.~Peterson\altaffilmark{1},
         Matthias~Dietrich\altaffilmark{1},
         Marianne~Vestergaard\altaffilmark{2}, and
         Misty~C.~Bentz\altaffilmark{1,3} }
         
\altaffiltext{1}{Department of Astronomy, 
		The Ohio State University, 
		140 West 18th Avenue, 
		Columbus, OH 43210; 
		denney, bentz, dietrich, 
		peterson@astronomy.ohio-state.edu}

\altaffiltext{2}{Department of Physics and Astronomy, 
		 Tufts University, 
		 Medford, MA 02155; 
		 m.vestergaard@tufts.edu}

\altaffiltext{3}{Present address: 
		 Dept. of Physics and Astronomy,
		 4129 Frederick Reines Hall,
		 University of California at Irvine,
		 Irvine, CA 92697-4575;
		 mbentz@uci.edu}

\begin{abstract}

We explore the nature of systematic errors that can arise in measurement
of black hole masses from single-epoch spectra of active galactic nuclei
(AGNs) by utilizing the many epochs available for NGC~5548 and
PG1229+204 from reverberation mapping databases.  In particular, we
examine systematics due to AGN variability, contamination due to
constant spectral components (i.e., narrow lines and host galaxy flux),
data quality (i.e., signal-to-noise ratio, $S/N$), and blending of
spectral features.  We investigate the effect that each of these
systematics has on the precision and accuracy of single-epoch masses
calculated from two commonly-used line-width measures by comparing these
results to recent reverberation mapping studies.  We calculate masses by
characterizing the broad \Hbeta\ emission line by both the full width at
half maximum and the line dispersion and demonstrate the importance of
removing narrow emission-line components and host starlight.  We find
that the reliability of line width measurements rapidly decreases for
$S/N$ lower than $\sim$ 10 to 20 (per pixel) and that fitting the line
profiles instead of direct measurement of the data does not mitigate
this problem but can, in fact, introduce systematic errors.  We also
conclude that a full spectral decomposition to deblend the AGN and
galaxy spectral features is unnecessary except to judge the contribution
of the host galaxy to the luminosity and to deblend any emission lines
that may inhibit accurate line width measurements.  Finally, we present
an error budget which summarizes the minimum observable uncertainties as
well as the amount of additional scatter and/or systematic offset that
can be expected from the individual sources of error investigated.  In
particular, we find that the minimum observable uncertainty in
single-epoch mass estimates due to variability is $\lesssim 0.1$ dex for
high $S/N$ ($\gtrsim 20$ pixel$^{-1}$) spectra.

\end{abstract}



\section{INTRODUCTION}
\label{S_Intro}

Understanding the demographics of supermassive black holes (BHs) is
imperative to expanding our understanding of the present state as well
as the cosmic evolution of galaxies.  In particular, links between BHs
and properties of host galaxies point to coevolution \citep{Kormendy95,
Ferrarese00, Gebhardt00a}. This is a surprising conclusion, given the
small sphere of influence of the central BH compared to the size of the
galaxy.  To trace the cosmic evolution of BHs, we must determine BH
masses as a function of cosmic time, thus requiring the measurement of
BH masses at large distances.  The direct methods (e.g. stellar and gas
dynamics, megamasers) that have succeeded for $\sim 30-40$ comparatively
local, mostly quiescent galaxies \citep[see review by][]{Ferrarese05}
fail at large distances because they require high angular resolution to
resolve motions within the radius of influence of the BH.  A solution to
this distance problem is to use AGNs as tracers of the BH population at
redshifts beyond the reach of the above mentioned methods.  AGNs are
luminous and easier to observe than quiescent galaxies at large
distances.  Most importantly, their masses can be determined by
reverberation mapping
\citep{Blandford82,Peterson93}, a method that does not depend on angular
resolution.

Masses have been measured for nearly $40$ active galaxies with
reverberation mapping (RM) methods \citep[see the recent compilation
by][]{Peterson04}. The results of these studies have led to the
identification of certain scaling relationships for AGNs. The
correlation between BH mass and bulge/spheroid stellar velocity
dispersion, i.e. the $M_{\rm BH}-\sigma_{\star}$ relation, for AGNs
\citep{Gebhardt00b, Ferrarese01, Onken04, Nelson04} is consistent with
that discovered for quiescent galaxies \citep{Ferrarese00, Gebhardt00a,
Tremaine02}.  In addition, and more relevant for this study, is the
correlation between the broad-line region (BLR) radius and the
luminosity of the AGN, i.e. the $R-L$ relation
\citep{Kaspi00,Kaspi05,Bentz06a, Bentz08}, which allows estimates of BH
masses from single-epoch (SE) spectra.  This relation affords great
economy of observing resources, allowing masses to be calculated for the
large number of AGNs/quasars with SE spectra obtained from surveys such
as the SDSS and AGES
\citep[e.g.,][]{Vestergaard02,Corbett03,Vestergaard04,Kollmeier06,Vestergaard08,JShen08,YShen08,Fine08}.

Measuring BH masses from single-epoch spectra presents itself as a
remarkably powerful tool for determining black hole masses at all
redshifts for potentially all spectroscopically observed quasars.
However, many effects limit the precision of these measurements, the
most important being how well the emission-line widths represent the
true motions of the BLR gas --- for example, if the BLR has a flattened
disk-like geometry, the unknown inclination of the system can result in
a huge uncertainty in the mass \citep{Collin06}.  For the moment,
however, if we set aside the issue of these calibration uncertainties
(i.e., the accuracy of the RM measurements themselves), the relevant
question becomes: how well can the SE mass measurements reproduce the RM
measurements?  Scaling relationships are being used with increased
frequency in the literature to indirectly measure black hole masses from
single-epoch spectra. Therefore, it is imperative that we understand the
systematic uncertainties introduced in measuring a SE black hole mass.
In particular, there are four systematics we will discuss that clearly
affect how well SE mass estimates reproduce the RM measurements of BH
masses.

\begin{description}
\item[{\bf 1. Variability:}] The
most inherent and unavoidable systematic in measuring masses of AGNs is
intrinsic variability that causes the luminosity, line widths, and
reverberation lag to change with time.  The variable luminosity leads to
variable BLR radius determinations when the $R-L$ relation is used, so we
are therefore likely to measure different SE masses for different
epochs.  In the case of AGNs for which multiple measurements of line
widths and radii are available from reverberation studies, the
relationship between line width and BLR radius (i.e. reverberation lag)
is consistent with a virial relationship, $\Delta V \propto R^{-1/2}$
\citep{Peterson99b,Peterson00a,Onken02,Kollatschny03}, as expected if
the BLR dynamics are dominated by gravity. This relationship also seems
to hold at least approximately for individual emission lines measured at
different times \citep[e.g.][]{Peterson04}: the size of the BLR as
measured in a particular emission line scales with luminosity
approximately as $R \propto L^{1/2}$ so we would expect that the line
width would correspondingly decrease as $\Delta V \propto L^{-1/4}$ in
order to preserve the virial relationship. Evidence to date suggests
that the central mass deduced from reverberation experiments at
different epochs is constant or, at worst, a weak function of luminosity
\citep{Collin06}. This is itself quite remarkable since we are
characterizing a region that is undoubtedly rather complex
\citep[cf.][]{Elvis00} by two quantities, the average time for response
of an emission line to continuum variations and the emission-line width.
In a previous investigation of variability on SE mass measurements using
\civ\,$\lambda1549$ emission, \citet{Wilhite07} show that the
distribution of fractional change in $M_{\rm BH}$ between epochs for
several hundred SDSS quasars has a dispersion of $\sim 0.3$.  Only part
of this dispersion can be accounted for by random measurement
errors.  Similarly, \citet{Woo07} estimate the uncertainty in SE mass
measurements based solely on propagating the variability in the
measurement of the H$\beta$ FWHM.  They demonstrate that the uncertainty
is roughly $30\%$.  However, the $S/N$ of their data was low ($\sim
10-15\, {\rm pixel}^{-1}$), and they attribute a large fraction of their
measured uncertainty to random measurement errors in the line width.
Here we will investigate the effect of variability by determining the
consistency of masses based on the two observables from optical SE
spectra: the monochromatic luminosity at $5100$\AA\ and the \Hbeta\ line
width.  We will use several hundred spectra of the Type~1 AGN NGC~5548
and a smaller sample of spectra of the Palomar Green (PG) quasar
PG1229+204.

\item[{\bf 2. Contamination by Constant Components:}] The variable AGN
spectrum is contaminated by relatively constant components.  These
include narrow emission lines and host galaxy starlight.  As the AGN
luminosity varies, so does the relative contributions to the observed
spectrum from these sources.  We will determine how the SE mass
measurements are affected by these non-variable features in the
spectrum.  In particular, we will examine changes in the precision and
accuracy of the masses when these contaminating features remain in the
spectrum, compared to when their contributions are subtracted before
luminosities and line widths are measured.

\item[{\bf 3. Signal-to-Noise Ratio:}] Accurately measuring the
spectroscopic properties needed for calculating the SE mass is highly
dependent on the quality of the spectra.  This is particularly true for
measuring emission-line widths.  Certainly not all spectra used for such
calculations in the literature are of comparable quality.  Therefore, we
will demonstrate how changes in the signal-to-noise ratio ($S/N$) of the
data affect SE mass measurements.  To make this comparison, we will
artificially degrade the $S/N$ of our sample to various levels and
compare the resulting masses.  In addition, it is common practice
\citep[e.g.,][]{McLure04, Woo07, YShen08, McGill08} to fit functions to
emission-line profiles in data with comparatively low $S/N$ in the hopes
of yielding more accurate line-width measurements.  Here we test
the usefulness of this practice by calculating and comparing SE masses
using line widths measured directly from the data to those using line
widths measured from fits to the line profiles of the original and
$S/N$-degraded spectra.

\item[{\bf 4. Blending:}] The optical region of broad-line AGNs is often
characterized by blending from many broad emission features as well as
contributions from the host galaxy starlight and AGN thermal emission.
Therefore, detailed modeling and decomposition of a spectrum into
individual spectral components is useful for isolating the features
required for accurately measuring SE masses.  However, this process is
rather time consuming as well as non-unique, since it requires
assumptions about the types of templates to fit and the relative
contributions to fit for each SE spectrum.  Instead, to make SE mass
measurements for a large number of AGNs, it is expedient to use simple
algorithms or prescriptions for these measurements.  There is concern,
however, that AGN emission-line blending and host galaxy features can
affect the accuracy of the line width measurements that utilize these
simple prescriptions.  With these considerations in mind, we will
compare SE mass measurements made from measuring spectral properties
using a simple prescription for local continuum fitting and subtraction
versus detailed modeling and decomposition of the optical region to
remove any extraneous components.
\end{description}

This is not a comprehensive list of systematics, but these particular
issues have a common element: all can be addressed empirically using a
large number of SE spectra of a single variable source.  The use of a
single source (actually, two single sources) is what sets this study
apart from past investigations, particularly on the point of
understanding the effect of variability on SE masses.  This is an
important distinction, given that \citet{Kelly07} show that an intrinsic
correlation between $M_{\rm BH}$ and $L$ that is statistically
independent of the $R-L$ relationship \citep[supported by,
e.g.][]{Corbett03,Netzer03,Peterson04} can lead to an artificially
broadened SE mass distribution when it is composed of masses from
multiple sources.  They suggest that because of this intrinsic $M_{\rm
BH} - L$ relation, using the luminosity simply as a proxy for the BLR
radius may cause additional scatter in the mass estimates because
additional information about the BH mass that may be contained in $L$ is
ignored.  By utilizing many epochs from the same object, however, the
effect on SE masses due strictly to variability can be isolated, while
the broadening caused by a possible $M_{\rm BH} - L$ correlation is
avoided, because we are dealing with a single black hole mass.
Therefore, any additional information about $M_{\rm BH}$ contained in
the luminosity could only affect the overall accuracy of our SE
measurements, not the scatter in our mass distributions due to
variability.

For each of the potential sources of uncertainty listed above, we
consider the effect on the precision of the SE mass estimates, which is
determined from the dispersion of these masses about the mean sample
value.  We will also consider the accuracy of the SE measurements, which
we define as the systematic offset between the distribution average and
a single mass based on reverberation mapping results for the same
sample.  We are not, however, addressing the accuracy of the
reverberation mapping masses themselves.  Better understanding and
quantifying the systematic uncertainties and zero-point calibration of
the reverberation mapping mass scale is an important but difficult
endeavor and will therefore be the focus of future work.  Because the
focus of this paper is not the accuracy of the RM measurements but
instead on the reproducibility of these values by SE measurements, we
will work only with the virial product, given by

\begin{equation}
M_{\rm vir} = \frac{c \tau (\Delta V)^2}{G},
\end{equation}

\noindent where $\tau$ is the measured time delay between the continuum
and broad emission-line variations (so that $c\tau$ is the effective BLR
radius) and $\Delta V$ is the velocity dispersion of the BLR gas.  Here,
we measure the velocity dispersion from the width of the broad H$\beta$
emission line.  By dealing simply with the virial product, or virial
mass, we bypass the zero-point calibration issue with the actual black
hole mass\footnote{This mass can be determined by scaling $M_{\rm vir}$
by a factor, $f$, which accounts for the unknown BLR geometry and
kinematics \citep[e.g.,][]{Onken04,Collin06,Labita06,Decarli08}.},
$M_{\rm BH}$.  In addition, we will consider without prejudice the two
common measures for characterizing line widths: the full width at half
maximum (FWHM) and the line dispersion, or second moment of the line
profile, $\sigma_{\rm line}$.

\section{Data and Analysis}
\label{S_Data_analysis}

\subsection{NGC 5548 Spectra}
\label{S_ngc5548}

The extensive, multi-decade monitoring of the Seyfert 1 galaxy NGC~5548
has led to one of the largest collections of observations of any single
AGN.  The size of this data set alone makes this object an obvious
choice for studying SE mass measurements.  The spectra of NGC 5548 for
this paper were selected from the International AGN Watch public
archives\footnote{http://www.astronomy.ohio-state.edu/$\sim$agnwatch/}.
With spectral quality in mind, we choose several subsets of the
available $1494$ spectra for the different analyses within this paper.
In our analysis of AGN variability and the effects due to the constant
spectral components in \S\S \ref{S_Res_variability} and
\ref{S_Res_const_components}, we use a total of $370$ spectra, including
the ``Revised selected optical spectra (1989-1996)'' covering years
$1-5$ \citep{Wanders&Peterson96}, as well as the remaining spectra from
the $1.8$m Perkins Telescope at Lowell Observatory covering years $6-10$
\citep{Peterson99a,Peterson02}.  This subset of data represents a nearly
homogeneous set of high-quality spectra that is centered on the
H$\beta\,\lambda 4861$ region of the optical spectrum.  We then use a
smaller subset of this NGC 5548 data set for the $S/N$ analysis in \S
\ref{S_Res_SN}, separating from the 370 spectra only those 270 observations
made with the Perkins Telescope.  These 270 spectra were all obtained
with the same instrument and instrumental setup, which kept properties
such as the entrance aperture, spectral resolution, and wavelength range
nearly constant for all observations.  This sample allows us to target
the systematic errors due to changes in $S/N$ rather than additional
observational systematics.  For the analysis of spectral component
blending covered in \S \ref{S_Res_decomp}, we focus on a set of 33
spectra from years $6-13$ of the AGN Watch campaign
\citep{Peterson99a,Peterson02} observed with the $3.0$m Shane Telescope at
Lick Observatory.  These spectra have full optical wavelength coverage
spanning rest frame $\sim 3000-7000$ \AA.  Utilizing this wide spectral
coverage, we perform full AGN-host spectral decompositions using two
independent methods to better judge the effects of blending.

Each of the above data sets have been internally flux calibrated to the
[O\,{\sc iii}]\,$\lambda 5007$ line flux in the mean spectrum using a
$\chi^{2}$ minimization algorithm developed by \citet{vanGroningen92}.
In this method the narrow emission-line flux can be taken as constant,
since these lines arise in an extended, low density region and are thus
unaffected by short timescale variations in the ionizing continuum flux.
Following this internal flux calibration, all subsets were scaled to the
absolute [O\,{\sc iii}]\,$\lambda 5007$ line flux of $5.58 \times
10^{-13}$ erg s$^{-1}$ cm$^{-2}$ \citep{Peterson91}.

\subsection{PG1229+204 Spectra}
\label{S_pg1229}

The PG quasar PG1229+204 (hereafter PG1229) from the Bright Quasar
Survey was chosen as an additional object for this study because it is
also a Type~1 AGN with a reverberation mapping mass measurement.  In
contrast to NGC~5548, however, it is a higher luminosity source where
neglecting the host galaxy and narrow lines is less likely to interfere
with the SE mass measurement.  In addition, results for this object will
allow for a more meaningful comparison (than the low-luminosity Seyfert
1 NGC~5548) with other quasars for which the $R-L$ scaling method is
more relevant.  The 32 optical spectra in our sample were originally
published along with reverberation mapping results for several PG
quasars by \citet{Kaspi00} and reanalyzed by \citet{Peterson04}.  Here,
we are again interested in the H$\beta$ region of the optical spectrum.
The absolute spectral fluxes of these data were calibrated externally
with comparison stars in the same field as the object \citep[for further
details see][]{Kaspi00}.

\subsection{Methodology for Measuring Virial Masses}
\label{S_measuringVPs}

\subsubsection{Virial Masses from Single-Epoch Spectra}
\label{S_SEVPmeasure}

The virial mass can be measured from a single optical spectrum by using
the width of the broad H$\beta$ emission line as a measure of the BLR
velocity dispersion and $\lambda L_{\lambda}$ at $\lambda=5100$ \AA\ in
the rest frame as a proxy for the BLR radius, $c\tau$, through the use
of the $R-L$ scaling relation \citep[e.g.][]{Kaspi00, Bentz06a,
Bentz08}.  We use the $R-L$ relation of \citet{Bentz08} because it
includes the most current reverberation mapping results and luminosities
that have been corrected for host galaxy starlight contamination.  Using
this form of the $R-L$ scaling relation and the virial mass formula
given by equation 1 (i.e., excluding any assumptions about the scale
factor, $f$), the SE virial mass is given by

\begin{equation}
\rm log \left(\frac{M_{\rm SE}}{\rm M_{\odot}}\right)=-22.0+0.519\,\rm log\left(\frac{\lambda L_{5100}}{\rm erg\,s^{-1}}\right)+2\,\rm log\left(\frac{V_{\rm H\beta}}{\rm km\,s^{-1}}\right),
\end{equation}

\noindent where $\lambda L_{5100}$ is the luminosity at rest frame
wavelength $5100$\,\AA, and $V_{\rm H\beta}$ is the line width of the
broad H$\beta$ emission line.  SE masses have been calculated in the
literature using various combinations of line widths and luminosity
measurements \citep[for examples and comparisons, see][]{McGill08}.
Therefore, we calculate eight virial masses for each SE spectrum using
different combinations of line width and luminosity measurements.
Through comparisons of these different mass estimates, we observe how
the systematics listed above affect the resulting SE masses in relation
to each of the spectral properties that we isolate in our calculation.

For the investigations of AGN variability, constant components, and
$S/N$ in \S\S \ref{S_Res_variability} -- \ref{S_Res_SN}, the continuum
flux density is taken as the average between observed-frame wavelengths
$5170$\,\AA\ and $5200$\,\AA\ for NGC~5548 and between $5412$\,\AA\ and
$5456$\,\AA\ for PG1229. These flux densities were corrected for
Galactic extinction, and then luminosity distances were calculated
assuming the following cosmological parameters: $\Omega_{m}=0.3$,
$\Omega_{\Lambda}=0.70$, and $H_0 = 70$ km sec$^{-1}$ Mpc$^{-1}$.
Luminosities for each spectrum were calculated both from the measured
continuum flux density and from the host galaxy-subtracted flux density.
For NGC5548 spectra the AGN continuum was then subtracted from each
spectrum based on a linearly interpolated fit between two local
continuum regions: one blueward of \Hbeta\ over the observed-frame range
$4825 - 4840$\,\AA\ and one redward of [O\,{\sc iii}]\,$\lambda 5007$
over the range $5170-5200$\,\AA.  Similarly, local continuum regions
were defined for PG1229 over the ranges $5063-5073$\,\AA\ and
$5412-5456$\,\AA.

Following continuum subtraction, H$\beta$ line widths are measured from
each spectrum within the following observed-frame wavelength ranges.
For the majority of the NGC~5548 data, H$\beta$ is defined over the
wavelength range $4845-5018$\,\AA\ for spectra with narrow line
components still present but was extended to the range $4845-5036$\,\AA\
for spectra from which we have removed the narrow lines because it is
often clear that the H$\beta$ profile extends under the [O\,{\sc
iii}]\,$\lambda 4959$ emission line.  However, during year 4 of the AGN
Watch campaign (JD2448636--JD2448898), NGC~5548 was in an extremely low
luminosity state, thus necessitating a different choice for the \Hbeta\
line boundaries and local continuum region blueward of this line.  For
spectra observed this year, we extended the boundaries of \Hbeta\ to
$4810-5135$\AA\ and defined the local continuum region blueward of
\Hbeta\ to be between $4782-4795$\AA.  For the PG1229 data set, H$\beta$
is defined over the range $5075-5248$\,\AA\ or $5075-5310$\,\AA\ for
spectra with and without narrow lines, respectively.  We measure the
line dispersion from the blue side of the broad H$\beta$ line, \sigbl,
assuming a symmetric profile about the line center.  This is done to
avoid residuals from the \ob\ narrow emission-line subtraction as well
as possible Fe\,{\sc ii} contamination commonly present on the red side
of the profile.  The FWHM is measured from the full line profiles
described above.  The exact procedures used for measuring these line
widths follow those of \citet{Peterson04}.  We measure the line widths
directly from the data, except in one subsection of the $S/N$ analysis
(\S \ref{S_VPs_from_fitsSN}), where line widths are measured from
Gauss-Hermite polynomial fits to the \Hbeta\ line profile.  Figure
\ref{fig:VandLvsTime} shows the host-subtracted luminosity and line
widths measured from the $370$ narrow-line subtracted SE spectra of
NGC~5548; the left panels show these observables as a function of time,
and the right panels show corresponding distributions, with the mean and
dispersions listed.  The dispersions in these quantities are non-random
and due primarily to the intrinsic variability of the AGN but also
include small random measurement uncertainties.

\subsubsection{Reverberation Virial Masses}
\label{S_RMmeasure}

To effect the most meaningful comparison with the single-epoch masses,
we calculate reverberation-based virial products, $M_{\rm vir}$, for
each data set.  Using radii from reverberation mapping leads to masses
that are independent of the uncertainties introduced in obtaining SE
radii measurements (i.e., AGN variability and calibration uncertainties
in the $R-L$ scaling relationship).  We use the reverberation radii of
\citet{Peterson04} that are derived from the rest-frame lag, $\tau_{\rm
cent}$, the centroid of the cross-correlation function.

We characterize the BLR velocity dispersion by both FWHM and \sigbl\ of
the broad H$\beta$ emission line.  Line widths are measured in the mean
spectrum for each observing season of NGC~5548 (years 1--13) created
from the sample of SE spectra used in each analysis and the full PG1229
data set after removal of the narrow-line components.  We use the same
methods and line boundaries as were used for the SE spectra.  Here, we
measure line widths in the mean spectrum \citep{Collin06} rather than
the rms spectrum \citep{Peterson04} because there is no analog for the
rms spectrum for a SE spectrum.  Instead, by using the mean spectrum, we
are still measuring the approximate mean BLR velocity
dispersion\footnote{The main justification for using the rms spectrum is
that only the portions of the line profile varying in response to the
ionizing continuum contribute to the rms spectrum.  See \citet{Collin06}
for a discussion.} yet retain a comparable line profile to a
single-epoch spectrum to use for a direct comparison.  Uncertainties in
these line width measurements are determined with the bootstrap method
of \citet{Peterson04}. We then combine the reverberation radii for each
year of the NGC~5548 sample and the single radius for PG1229 with the
corresponding values of each line width measurement to calculate two
sets of RM virial products for each data set: one using FWHM and one
using \sigbl.  Weighted mean virial products are then calculated for the
NGC~5548 data sets spanning multiple years: $1-10$ for the variability,
constant component, and $S/N$ analyses in
\S\S \ref{S_Res_variability} -- \ref{S_Res_SN} and $6-13$ for the
blending analysis in \S \ref{S_Res_decomp}, providing two final
reverberation virial masses for each data set: one using \sigbl\ and one
using FWHM.

\subsubsection{Comparisons:  Measuring Precision and Accuracy}
\label{S_VPcomparisons}

We measure the precision of SE virial masses by creating distributions
of the SE virial masses calculated for each data set as described above.
The dispersion, $\sigma_{\rm SE}$, of these distributions serves as a
measure of the precision of the SE masses.  It gives an indication of
how well multiple SE spectra can reproduce a single, mean mass, $\langle
{\rm log}\,M_{\rm SE} \rangle$.  In addition, the accuracy of the SE
masses can be gauged by measuring the systematic offset of this mean
mass from the corresponding reverberation virial mass determined for a
given data set.  We define this offset as $\langle \Delta {\rm
log}\,M \rangle = \langle {\rm log}\,M_{SE} \rangle - {\rm
log}\,M_{\rm vir}$ and calculate a value for each SE mass distribution.

\subsection{Evaluation of Constant Components}
\label{S_Eval_const_components}

A copy of each flux calibrated spectrum was made, and the narrow emission
lines were removed from this copy to allow for the calculation and
comparison of virial products from spectra with and without narrow lines
present.  Narrow H$\beta\,\lambda 4861$ and the [O\,{\sc iii}]\,$\lambda
\lambda 4959, 5007$ lines were removed by first creating a template
narrow line from the [O\,{\sc iii}]\,$\lambda 5007$ line in the mean
spectrum from each data set.  This template was then scaled in flux to
match and remove the [O\,{\sc iii}]\,$\lambda 4959$ line and narrow
component of H$\beta$ \citep{Peterson04}.

Host galaxy starlight contributions to the flux were determined for the
various extraction apertures of all NGC~5548 and PG1229 spectra using
the method of \citet{Bentz08} and observations of both galaxies with
the High Resolution Channel of the Advanced Camera for Surveys on the
{\it Hubble Space Telescope}.  Luminosities (and subsequently SE virial
masses) were calculated for every spectrum with and without the presence
of this constant continuum component.

\subsection{Spectral Decomposition: Deblending the Spectral Features}
\label{S_Spect_decomp}

The rather simple approach used above to measure line widths and to
account for host galaxy contamination using local continuum fitting
techniques fails to address certain spectral features or components that
may systematically affect our SE virial mass estimates.  First, the
global AGN continuum is power-law shaped, rather than linear, as we fit
above.  This may lead to small uncertainties in our continuum
subtraction, possibly even over the small wavelength range used here.
Second, blended Fe\,{\sc ii} emission exists throughout the optical
spectrum.  If strong, this emission could complicate our definitions of
local continua on either side of the \Hbeta, \ob\ region, potentially
adding flux both to these continuum regions and to H$\beta$ itself.
Third, the red wing of broad \HeII\ emission may be blended with the
blue wing of H$\beta$.  This could also contaminate the local continuum
region defined between these two lines as well as the line width
measurement.  Fourth, the underlying galaxy spectrum has structure that
may be imprinted on the broad line profiles if not removed accurately.

Therefore, we undertake full spectral decompositions of a selection of
NGC~5548 spectra.  Our goal is to determine if the (potentially over-)
simplified local continuum-fitting prescription to account for the
underlying host galaxy and additional AGN emission skew the SE virial
mass results in a significant, yet correctable manner.  The data set
used in this section consists of the 33 Shane Telescope spectra from the
AGN Watch sample described above.  Because spectral decomposition gives
model-dependent, non-unique solutions, we compare results based on two
independent methods utilizing multi-component fits to account for
contributions from the host galaxy, AGN continuum, and spectral emission
lines.

\subsubsection{Method A}
\label{S_describe_methodA}

Decomposition method A \citep[e.g.,][]{Wills85} assumes that the observed
spectra can be described as a superposition of five components
\citep[see][for more details]{Dietrich02,Dietrich05}:
\begin{enumerate}

\item An AGN power law continuum ($F_\nu \sim \nu ^{\alpha}$).  

\item A host galaxy spectrum \citep{Kinney96}. 

\item A pseudo-continuum due to merging \feii\ emission blends. 

\item Balmer continuum emission \citep{Grandi82}.

\item An emission spectrum of individual emission lines, such as
H$\alpha$, [\nii]$\lambda \lambda 6548,6583$, \hei\,$\lambda 5876$,
H$\beta$, [\oiii]$\lambda \lambda 4959,5007$, \heii\,$\lambda 4686$.
\end{enumerate}

The first four components are simultaneously fit to each single-epoch
spectrum, minimizing the $\chi^2$ of the fit. We tested several
different host galaxy templates (elliptical galaxies, S0, and spiral Sa
and Sb galaxies). The best results were obtained using a scaled spectrum
of the E0 galaxy NGC\,1407, which is quite appropriate for the bulge of
NGC\,5548.  From this template we measure an average host starlight
contribution of $F_{\rm gal}(5100$\,\AA$) = (4.16\pm0.84)\times
10^{-15}$ \Flamunit\ for this sample.  This is highly consistent with
the value of $F_{\rm gal}(5100$\,\AA$) = (4.45\pm0.37)\times 10^{-15}$
\Flamunit\ derived with the \citet{Bentz08} procedure for the observed
aperture of ($4\arcsec \times 10\arcsec$) for this data. To account for
the \feii\ emission, we use the rest-frame optical template covering
$4250-7000$ \AA\ based on observations of I\,Zw1 by
\citet{Boroson92}.  The width of the \feii\ emission template was on
average FWHM\,=\,$1160\pm34$ km\,s$^{-1}$. For the Balmer continuum
emission, we found that the best fit was obtained for $T_e = 15,000$\,K,
$n_e = 10^8$\,cm$^{-3}$, and optically thick conditions.  

The best fits of these components, including the power-law fit to
account for AGN continuum emission, are subtracted from each spectrum,
leaving the AGN emission-line spectrum intact.  Narrow emission-line
components were then subtracted by creating a template narrow line from
a two-component Gaussian fit to the [\oiii]\,$\lambda 5007$ narrow line
and then scaling it to each individual narrow line to be subtracted
based on standard emission line ratios.  Figure \ref{fig:MDdecompFig}
illustrates the different fit components and residuals for a typical
spectrum of NGC\,5548.  Overall, the spectrum is quite well
reconstructed.  However, it can be seen that around $\lambda \simeq
5200$\,\AA\ to $\lambda \simeq 5800$\,\AA\ the flux level is
overestimated. This might indicate that it is necessary to include an
additional component due to Paschen continuum emission, as suggested by
\citet{Grandi82} and more recently by \citet{Korista01}.  This may, in turn,
result in the selection of a less red host galaxy spectrum but
potentially a better overall fit \citep{Vestergaard08}. This component
was not included here, however, because the actual strength of the
Paschen continuum emission is not yet well constrained and will
therefore be the topic of future work in this area.

\subsubsection {Method B}
\label{S_describe_methodB}

This method first corrects the spectra for Galactic reddening using the
extinction maps of \citet{Schlegel98} and the reddening curve of
\citet{ODonnell94} with $E(B-V) = 0.0392$.  The continuum and emission
lines were modeled separately.  The continuum components include the
following:
\begin{enumerate}

\item A nuclear power-law continuum.

\item The \feii{} and \feiii{} blends that form a
``pseudo-continuum'' across much of the UV-optical range.  Template
modeling is the only way, at present, to provide a reasonable iron
emission model for subtraction \citep[see, e.g.,][and references
therein]{Vestergaard01, Veron-Cetty04}.  We used the optical iron
template of \citet{Veron-Cetty04}, varying only the strength of the
template and the line widths.

\item The Balmer continuum was modeled using the prescription of
\citet{Grandi82} and \citet{Dietrich02}, with an adopted electron
temperature of $10,000$ K and an optical depth at the Balmer edge of
$1.0$.  These Balmer continuum parameters typically give good matches to
quasar spectra \citep{Vestergaard08}.  We note that here, too, no
attempt was made to model the Paschen continuum as it is poorly
constrained.

\item The underlying host-galaxy spectrum was modeled using the
stellar population model templates of \citet{Bruzual03}. The best-fit
model was a single elliptical galaxy template with stellar ages of 10
Gyr.  This model seems to slightly underestimate the stellar emission
strength in NGC 5548 longward of H$\alpha$ (e.g., $\gtrsim 7000$\AA),
but preliminary fits (not included) seem to indicate the overall fit is
better with the inclusion of the Paschen continuum.
\end{enumerate}

The individual continuum model components were varied to provide the
optimum match to the observed spectrum using Levenberg-Marquardt
least-squares fitting and optimization. Based on the host galaxy
template fits for this decomposition method, we measure a host starlight
contribution of $F_{\rm gal}(5100$\,\AA$) = (7.05\pm1.28)\times
10^{-15}$ \Flamunit, larger than found by \citet{Bentz08} and the
method A value, but marginally consistent once these other values take
Galactic reddening into account. The best fit continuum components were
then subtracted from the spectrum, and the remaining emission-line
spectrum was modeled with Gaussian functions using the same optimization
routine as for the continuum.  A single Gaussian profile, whose width
was allowed to vary up to 600 km s$^{-1}$ was used for each of the
narrow emission lines, but the same width was used for all narrow
lines. The strength of the \ob\ doublet lines was constrained to the 1:3
ratio set by atomic physics.  Figure \ref{fig:MVdecompFig} shows the
individual and combined components fit to the same NGC~5548 spectrum as
in Figure \ref{fig:MDdecompFig}, as well as the residual spectrum after
subtraction of both continuum components and the narrow emission lines.

\subsubsection{Methodical Differences and Mass Calculations}
\label{S_method_diff_andVPs}

Overall, the two methods for fitting the individual AGN spectral
components agree quite well.  However, there are two differences worth
noting.  First, each method uses a different optical \feii\ emission
line template.  The \feii\ template used by method B
\citep{Veron-Cetty04} includes narrow line region contributions to the
emission.  In general, both templates are similar, but they differ in
detail at around $\lambda \simeq 5000$\,\AA\ and $\lambda \gtsim
6400$\,\AA. However, the strength of the optical \feii\ emission in
NGC\,5548 is quite weak \citep{Vestergaard05}, and the width of the
\feii\ emission is expected to be broad.  Therefore, the choice of the
\feii\ emission template has little impact on the results in this
case.  Second, the modeling of the narrow emission lines in the \Hbeta\
and \ob\ region with the single Gaussian component of method B sometimes
leaves some residuals around the \ob\ lines.  This typically happens
when excess emission appears in the red wing of \Hbeta\ which cannot be
fully accounted for with only Gaussian components.  Although these
residuals do not account for a significant amount of flux, the
two-component Gaussian fit to the narrow lines used by method A tends to
better minimize these residuals.

Line widths are measured for \Hbeta\ from all epochs following analysis
from both spectral decomposition methods as well as the local continuum
fitting method.  For this particular sample, the local continuum was
defined between two continuum windows over the rest-frame wavelength
ranges $4730-4745$\AA\ and $5090-5110$\AA, and line widths were measured
in all spectra over the rest-frame wavelength range $4747-4931$\AA, as
determined from the mean spectrum.  For the local continuum-fitted
spectra, $L_{5100}$ was calculated from the average continuum flux
density over the rest-frame wavelength range $5090-5110$\AA\ after
correcting for host galaxy starlight.  For decomposition methods A and
B, $L_{5100}$ was taken to be the value of the power-law fit to each SE
spectrum at rest-frame $5100$\AA.  SE virial masses were then calculated
with equation 2 for line widths measured with both \sigbl\ and FWHM.

For comparison to the SE mass distributions of each of the three data
analysis methods, reverberation virial masses were calculated with each
of the line width measures, FWHM and \sigbl, similar to the previous
NGC~5548 data sets spanning multiple years.  The weighted mean RM virial
mass for each analysis method (covering yrs 6--11 and 13 for this data
set) was calculated by averaging the yearly RM virial masses calculated
by combining line widths measured from the mean spectrum created from SE
spectra spanning a single observing season and the BLR radius from the
corresponding season as determined with reverberation mapping \citep[for
results from individual years, see][]{Peterson04}.

\section{Analysis and Results}
\label{S_Results}

\subsection{Effects of Variability}
\label{S_Res_variability}

To investigate systematics associated strictly with AGN variability in
SE virial products (VPs), we first remove the contaminating constant
spectral components (i.e., narrow lines and host galaxy flux) as
described in \S \ref{S_Eval_const_components}.  Figure
\ref{fig:SEVPdistr_var} shows virial mass distributions created from all
370 spectra of NGC~5548 (Fig. \ref{fig:SEVPdistr_var}a) and 32 spectra
of PG1229 (Fig. \ref{fig:SEVPdistr_var}b).  Results are shown for both
line width measures, \sigbl\ and FWHM (left and right panels,
respectively), for both objects.  For each distribution we focus on the
dispersion (i.e., precision) and the mean offset (i.e., accuracy),
$\langle \Delta {\rm log} M \rangle $, from the reverberation result, as
given in Table~1. Column 1 gives the object name, column 2 shows the
sample size, column 3 lists the reverberation virial product and
associated uncertainties when calculated with \sigbl, column 4 lists the
mean and standard deviation of the distribution utilizing \sigbl, column
5 gives the mean offset between the reverberation VP (Col. 3) and mean
of the SE distribution (Col. 4).  Columns 6, 7, and 8 are similar to
columns 3, 4, and 5, but for masses based on FWHM.

Figure \ref{fig:SEVPdistr_var} shows that the widths of all four
distributions are quite small.  The listed dispersions have not been
corrected for measurement uncertainties in line width and luminosity.
However, we have estimated the average measurement uncertainties for the
full set of NGC~5548 SE spectra to be $0.08$ dex in log($L$), $0.01$ dex
in log(\sigbl), and $0.03$ dex in log(FWHM). We can assume that these
measurement errors are independent of the dispersion due to variability
alone and that the distributions are close enough to Gaussian that we
can add independent errors in quadrature.  Therefore, we can correct the
observed dispersions in the SE mass distributions for NGC~5548 for the
contribution due to these measurement uncertainties.  Following this
correction, the uncorrected dispersions listed in Figure
\ref{fig:SEVPdistr_var}a and Table 1 can be reduced to $0.11$ dex for
masses based on \sigbl\ and $0.14$ dex for masses based on FWHM.  The
narrowness of these distributions indicates that the scatter in SE
masses due to intrinsic variability is remarkably small.  This is
particularly true for PG1229, for which $\sigma_{\rm SE} \approx 0.05$
dex.  Granted, PG1229 is less variable, but with a scatter of only
$0.11-0.14$ dex, the uncertainty in $M_{\rm SE}$ due to variability for
NGC~5548 is not large either.

The precision and accuracy in the $M_{\rm SE}$ measurements seem only
weakly dependent on whether \sigbl\ or FWHM is used as the line-width
measure.  For both AGNs, the scatter is apparently minimized and the
accuracy (given by $\langle \Delta {\rm log} M \rangle$) maximized with
the use of \sigbl.  In terms of accuracy, $\langle \Delta {\rm log} M
\rangle$ should at least partially represent the displacement of the
particular AGN from the $R-L$ relation, regardless of which quantity is
used to characterize the line width.  Figure \ref{fig:rLrelation} shows
that the average luminosities of the NGC~5548 and PG1229 SE spectra
place them above the $R-L$ relation in $r$ by $\sim 0.13$ dex and $\sim
0.07$ dex, respectively, after accounting for host starlight
contributions (as was done here).  This explains why, for a given SE
luminosity, the resulting radius (and thus VP) is underestimated
compared to the reverberation results, confirmed by the negative
$\langle \Delta {\rm log} M \rangle$ values found in Table~1.  Since
this effect depends on luminosity alone, the masses calculated from both
line width measures should be affected equally.  However, masses
calculated from FWHM measurements result in larger $\langle \Delta {\rm
log} M \rangle$ values for both objects.  This additional component may
be related to the fact that our measurement uncertainties tend to be
larger for FWHM compared to \sigbl, or it may simply demonstrate one of
the limitations of measuring masses from SE spectra with FWHM.

The light curves of NGC 5548 span several years, much longer than the
reverberation time scale of tens of days. Indeed, the \Hbeta\ lag has
been measured year-to-year for over a dozen different years, and the lag
and the mean luminosity of the AGN are well-correlated on yearly
timescales, and as noted earlier, the reverberation-based mass is
approximately constant with perhaps a weak dependence on luminosity
\citep{Bentz07}. Given our goal of comparing SE predictions with
reverberation measurements, we have for NGC 5548 also computed the
difference between each SE virial product and the reverberation virial
product for the specific year in which the SE observation was made. We
show the distribution of these differences in Figure
\ref{fig:deltalogM}, which is rather narrower than the similar
distribution shown in Figure \ref{fig:SEVPdistr_var}a.  This illustrates
that masses from SE spectra seem to reproduce the reverberation mass
that would be measured at the same time quite accurately, to $\sim 25$\%
or so. However, there are longer term secular changes that occur, as
shown in the top panel of Figure 1, that add to the observed dispersion
due to variability resulting in the total width of the distributions
shown in Figure \ref{fig:SEVPdistr_var}.  Because of these secular
changes, even a reverberation-based mass measurement might change
slightly, say, over a dynamical time scale.

\subsection{Accounting for Constant Components}
\label{S_Res_const_components}

Failing to account for the constant spectral components in the AGN
spectrum (i.e., the narrow emission lines and host galaxy starlight)
affects both the precision and accuracy of the SE mass estimates.  We
examine the effect of neglecting each of these components individually
and then in combination for both NGC~5548 and PG1229.

\subsubsection{Effect of Starlight}
\label{S_Res_hosteffects}

First, we examine the consequence of failing to remove the host starlight
contribution to the continuum flux density.  We still subtract narrow
emission-line components, however.  Figure \ref{fig:SEVPnl_nogs} shows SE
virial mass distributions similar to those in Figure
\ref{fig:SEVPdistr_var}, but here the host starlight was not subtracted from
the luminosity before the SE masses were calculated.  In terms of
precision, the virial mass distributions in Figure \ref{fig:SEVPnl_nogs}
derived from non-host-corrected luminosities have equal or even slightly
smaller dispersions than their corrected counterparts
(Fig. \ref{fig:SEVPdistr_var}).  This occurs simply because subtraction
of the host starlight increases the relative amplitude of the AGN
continuum variations.  NGC~5548 has a relatively larger host galaxy
contribution and is therefore more susceptible to this effect than
PG1229.  The observable result is an overall increase in the dispersion
of the mass distribution and, in particular, the low-mass (i.e.,
low-luminosity state) wings of the $M_{\rm SE}$ distributions in Figure
\ref{fig:SEVPnl_nogs} are broadened compared to those in Figure
\ref{fig:SEVPdistr_var}.  Notably, over-subtracting the host galaxy flux
could also lead to similar observable consequences.  However, the tail
of the distribution appears to be nearly Gaussian, which argues against
any large error in the starlight flux estimate.  In contrast, this
broadening affect is not observed for PG1229.  This is expected because
PG1229 has a smaller host contribution to its total luminosity than
NGC~5548, and its luminosity varied less over the time period in which
it was observed.  Therefore, when we subtract a relatively smaller
constant host flux from a distribution of values with an initially
smaller luminosity dispersion, the effect on the SE mass distributions
is less significant.

Failing to account for host starlight imposes a shift to the entire SE
mass distribution.  Because the luminosity is larger when the host
contribution is not subtracted, a larger BLR radius is estimated with
the $R-L$ relation.  This, in turn, produces larger virial products and
affects the accuracy of the measurements.  Whereas Figure
\ref{fig:SEVPdistr_var} shows an average underestimation of the SE
masses compared to the reverberation results, Figure
\ref{fig:SEVPnl_nogs} shows that on average, the SE masses are
overestimated (i.e. positive $\langle \Delta {\rm log} M \rangle$
values), which is again explained by the locations of NGC~5548 and
PG1229 on the $R-L$ scaling relationship (Fig. \ref{fig:rLrelation}).
Without accounting for the host starlight, both objects lie below the
relation.  Therefore, the $R-L$ relation overestimates the radius of a
SE luminosity measurement that does not account for this contribution.
This effect can be seen by comparing the $\langle \Delta {\rm log} M
\rangle$ values in rows 2 and 5 of Table 2 with those of Table 1, which
are negative for in Table 1 but positive in Table 2\footnote{Results in
Table 2 are presented in a similar manner as Table 1, except columns
have been added to distinguish whether or not narrow emission-line
and/or host starlight contributions are present in the results.}.
Failing to account for the host contribution has roughly the same
overall effect on the precision and accuracy of SE mass distributions
regardless of whether \sigbl\ or FWHM is used for the calculation of
$M_{\rm SE}$ (a shift in $\langle \Delta {\rm log} M \rangle$ of $0.17$
dex for NGC~5548 and $0.12$ dex for PG1229 for both line width
measures), as expected since this contribution does not affect the line
width.  

Based on the results presented here, it is not completely clear that
subtracting the host contribution improves the overall accuracy of the
mass estimates. In fact, the SE masses of both NGC~5548 and PG1229
presented here are typically as accurate or more accurate (i.e., the
absolute value of $\langle \Delta {\rm log} M\rangle$ is smaller) when
the starlight contribution is not subtracted.  When considering the
physics of AGNs, however, the BLR radius should be correlated with only
the AGN luminosity, since the material in the BLR knows nothing of the
luminosity originating from galactic starlight.  Furthermore,
\citet{Bentz08} determine that calibrating the $R-L$ relation with luminosity
measurements that have been corrected for host starlight contamination
significantly reduces the scatter in the relationship and results in a
slope that is highly consistent with that predicted by simple
photoionization theory.  These considerations, in addition to our use of
the \citet{Bentz08} host starlight-corrected calibration of the $R-L$
relation for SE mass determinations, serve as motivation for removing
this contamination before the $R-L$ relation is used.  This evidence
suggests that the ambiguity between the theoretical expectation that
host-subtracted luminosities should yield more accurate masses and the
fact that the masses presented here are more accurate before host
starlight subtraction is simply because both NGC~5548 and PG1229 happen
to lie above the $R-L$ relation.  However, in a general statistical
sense, SE masses will be overestimated if host starlight contamination
is not taken into account before the $R-L$ relation is used to determine
BLR radii.  This is particularly true for lower-luminosity, Seyfert-type
galaxies that, in contrast to quasars, have larger relative host
starlight contributions to their measured luminosity.

\subsubsection{Effect of Narrow Lines}
\label{S_Res_NarLines}

The \Hbeta\ and \ob\ emission line profiles for NGC~5548 and PG1229 are
shown in Figure \ref{fig:meanspec}.  In NGC~5548 (left), the narrow line
typically increases the peak flux by $\sim 50\%$, compared to $\lesssim
10\%$ in PG1229 (right).  Given these relative contributions of
narrow-line fluxes (particularly in the case of NGC~5548), failing to
subtract the narrow line component from the broad emission line before
measuring the width can have a significant impact on the resulting mass
estimate.  To demonstrate this, Figure \ref{fig:SEVPdistr_wlgs} displays
SE mass distributions for NGC~5548 and PG1229; this time, however, we do
not subtract the narrow lines from the spectra before measuring line
widths, although we do subtract the host galaxy contribution.
Statistics for the scenarios shown in Figure \ref{fig:SEVPdistr_wlgs}
can be found in Table 2, rows 3 and 6.  Figure \ref{fig:SEVPdistr_wlgs}
clearly demonstrates that leaving the narrow lines present affects both
the precision and accuracy of the SE masses.

Failing to subtract the narrow lines tends to decrease the precision of
the SE mass estimates.  This is evident by an increase in the width of
the SE mass distributions and is particularly pronounced for NGC~5548
when characterizing the line with FWHM (by comparing
Fig. \ref{fig:SEVPdistr_wlgs}a with Fig. \ref{fig:SEVPdistr_var}a or
Fig. \ref{fig:SEVPnl_nogs}a, in which narrow lines were removed).  In
this case (Fig. \ref{fig:SEVPdistr_wlgs}a, right), the resulting width
of the VP distribution is a factor of three to four larger than if
\sigbl\ is used (Fig. \ref{fig:SEVPdistr_wlgs}a, left).  This effect of
the narrow lines on the precision is less apparent in PG1229 because the
narrow line constitutes only a small percentage of the \Hbeta\ line
flux.  However, it is still observed when the FWHM is used for measuring
the line widths (compare Fig. \ref{fig:SEVPdistr_wlgs}b, left, to
Fig. \ref{fig:SEVPdistr_var}b, left), since, to reiterate, this trend is
much more apparent for the FWHM.

From a physical standpoint, only BLR emission varies in response to the
ionizing continuum on reverberation timescales, so only the broad
emission component should be used for the virial mass calculation.
Because the square of the line width enters into the BH mass
calculation, relatively small changes in the line width can
significantly affect the mass estimate.  When the narrow-line component
is not subtracted, the line width and hence the black hole mass is
underestimated.  Figure \ref{fig:SEVPdistr_wlgs} shows evidence for this
in both NGC~5548 and PG1229.  As with the precision, this effect is much
stronger when the narrow component is a more prominent feature in the
emission-line profile, as is the case for NGC~5548 (refer back to
Fig. \ref{fig:meanspec}).  For obvious reasons, removing the narrow
lines is more important when the line width is measured with the FWHM
(right panels of Figs. \ref{fig:SEVPdistr_wlgs}a and
\ref{fig:SEVPdistr_wlgs}b); the very definition of the FWHM depends on
the peak flux, so if the narrow component is not subtracted, this peak
flux can be greatly overestimated.  An overestimation of the peak flux
results in an artificially small FWHM and, subsequently, a severely
underestimated mass.  NGC~5548 affords a useful case in point: the
masses calculated without removing narrow line components (Figure
\ref{fig:SEVPdistr_wlgs}a) are underestimated on average by a whole 
order of magnitude ($\langle \Delta {\rm log} M \rangle = -1.00$) when
line widths are measured from the FWHM (right panel).  In contrast, the
dependence of the line dispersion on the line center and peak flux is
relatively weak, affecting the accuracy of SE mass estimates by $\sim
0.1-0.2$ dex for NGC~5548 and by an insignificant amount for PG1229 (compare
$\langle \Delta {\rm log} M \rangle$ values for $M_{\rm SE}
\propto$ \sigbl\ from Table 2, Rows 3 and 6 to Table 1 values).
Regardless of the minimal effect when \sigbl\ is used, the evidence
presented here clearly indicates that the narrow line component should
be removed regardless of which prescription is used for measuring the
line width.

\subsubsection{Combined Effects of Starlight and Narrow Lines}

Figure \ref{fig:SEVP_nocorrections} shows mass distributions for both
NGC~5548 and PG1229 when neither of these constant components is removed
from the spectra.  Table 2 (Rows 1 and 4) displays the corresponding
statistics.  Generally, as expected, the precision and accuracy are
worse, or at least no better than when these constant components are
removed.  However, these two constant components act opposingly on the
mass: failing to remove the narrow lines tends to decrease mass
estimates, but failing to subtract host galaxy flux increases mass
estimates.  Therefore, these two effects can fortuitously cancel,
resulting in an apparently smaller dispersion and/or mean offset.  This
is the case for PG1229 when FWHM is used to measure the line width and
NGC~5548 when \sigbl\ is used.  The chance cancellation in these cases
should not distract from the otherwise well-supported conclusion that
both of these components should be removed to obtain the most accurate
and precise SE mass estimates.

\subsection{Systematic Effects due to \boldmath{$S/N$}}
\label{S_Res_SN}

Our goal here is to identify the point at which low $S/N$ begins to
compromise the precision and accuracy of SE mass determinations.  We
start with our most homogeneous data set, the 270 observations of
NGC~5548 from the Perkins Telescope.  Based on conclusions from previous
sections \S\S \ref{S_Res_variability} and \ref{S_Res_const_components},
only narrow-line-subtracted spectra that have been corrected for host
galaxy starlight are used.  The $S/N$ per pixel of the original spectra
ranges significantly, with a mean and standard deviation of $110 \pm
50$, as measured across the $5100$\AA\ continuum window given above.
Using the $S/N$ per pixel in the original spectra as a starting point,
we then increase the noise in each spectrum by applying a random
Gaussian deviate to the flux of each pixel across the whole spectrum.
The magnitude of the deviate is set to achieve degraded $S/N$ levels of
$\sim 20$, $\sim 10$, and $\sim 5$ across the $5100$\AA\ continuum
window.  Figure \ref{fig:SNdegrade} shows an example degradation for a
typical NGC~5548 spectrum.  Below, we discuss results for masses
calculated from line widths measured directly from the data as well as
from Gauss-Hermite fits to \Hbeta\ in the original and $S/N$ degraded
spectra.

\subsubsection {Direct Measurement of the Spectra}
\label{S_VPs_from_dataSN}

We measure line widths and luminosities directly from both the original
and $S/N$-degraded spectra and calculate virial masses.  The resulting
distributions are shown in Figure \ref{fig:SEVPdataSN} for both \sigbl\
(left) and FWHM (right).  Statistics describing the distributions of
$M_{\rm SE}$ are listed in Table 3 in a format similar to that of
previous tables.  Figure \ref{fig:SEVPdataSN} shows that the dispersions
of the distributions broaden as the $S/N$ of the spectra decreases.
Overall, low $S/N$ begins to negatively affect the precision of the
virial mass estimates at $S/N$ $\lesssim 10$ for \sigbl\ (see third
panel on left) and at $S/N$ $\lesssim 5$ for FWHM (see bottom panel on
right).  \citet{Wilhite07} find a similar result, with the widths of
their SE mass distributions increasing steadily with decreasing $S/N$.
However, measurements of \sigbl\ and FWHM are affected differently by
decreasing $S/N$ and will therefore be discussed separately.

Measurements of virial masses from \sigbl\ in low $S/N$ spectra
sacrifices both precision and accuracy primarily because the wings of
the broad line become lost in the noise and the line profile boundaries
cannot be accurately defined for cases where $S/N \lesssim 10$.  This
results in smaller effective line widths.  This effect decreases the
overall accuracy by shifting the whole distribution to artificially
smaller masses.  However, at these low $S/N$ limits (see bottom two
plots of Fig. \ref{fig:SEVPdataSN}, left), the distribution actually
becomes highly non-Gaussian in shape, resulting in a much peakier
distribution, nearly centered on the corresponding reverberation virial
product.  This implies that although the overall dispersion has
increased significantly (by nearly a factor of 2) and individual
measurements have the potential to be highly inaccurate, a typical
measurement will likely be more accurate with a much smaller uncertainty
than quoted through the overall distribution average.

Different systematics are introduced when using FWHM to characterize the
line width.  Because FWHM does not depend on the line wings, lower $S/N$
can be tolerated before the precision is significantly sacrificed.  When
$S/N$ is low enough to affect FWHM, the line width is generally
underestimated.  Several effects contribute to the difficulty in
defining FWHM in low $S/N$ data.  First, the peak flux may be
incorrectly attributed to the highest noise spike, resulting in an
overestimated maximum.  Second, the half-maximum may be difficult to
define because the continuum level cannot be accurately ascertained.
Third, the width may also be problematic to define because a noisy
profile could mean that the half-maximum flux value is shared by
multiple wavelength values.  These effects alter the precision at our
lowest degraded $S/N$ level ($\sim 5$).  However, they begin to affect
the accuracy of the measurement much earlier.  Progressively poorer
accuracy can be easily observed from the increasingly negative $\langle
\Delta {\rm log} M \rangle$ values in the distribution statistics given
in Table 3 for FWHM and/or by comparing the mean values of the
distributions in Figure \ref{fig:SEVPdataSN}, right.  Although higher
precision VP measurements can be made from lower $S/N$ data with the
FWHM than with \sigbl, there is a trade-off in accuracy.  For this
reason, we caution against measuring SE masses from spectra with $S/N$
lower than $\sim 20$ pixel$^{-1}$, regardless of the line-width
measurement method.

\subsubsection{Measurements from Gauss-Hermite Polynomial Fits}
\label{S_VPs_from_fitsSN}

Recent work has been published in which the emission line profiles are
fit with either Gaussian and/or Lorentzian profiles \citep[e.g.,][]{
McLure04, JShen08, YShen08} or Gauss-Hermite polynomials
\citep[e.g.,][]{Woo07, McGill08}.  SE virial masses are then calculated
with the line widths measured from these fits rather than directly from
the data in an attempt to mitigate the negative effects of low $S/N$ on
line-width determinations.  We test this technique by fitting a
sixth-order Gauss-Hermite polynomial to the narrow-line-subtracted
H$\beta$ profiles in the original and $S/N$-degraded spectra used above.
A linearly interpolated continuum defined by the same regions as above
was first subtracted from the spectra before the fits were made.  Our
Gauss-Hermite polynomials utilize the normalization of
\citet{vanderMarel93} and the functional forms of
e.g. \citet{Cappellari02}.  We then use the method of least-squares to
determine the best coefficients for the sixth-order polynomial fit.  The
thick black curves in Figure \ref{fig:SNdegrade} show an example of the
fits to the original and $S/N$-degraded forms of this typical NGC~5548
spectrum.  Both FWHM and \sigbl\ were measured from these fits with the
same methods described previously for the direct measurements and then
combined with host-corrected luminosities in order to calculate virial
masses for all SE spectra in this sample.  Figure \ref{fig:SEVPfitsSN}
shows the resulting distributions for the virial masses calculated using
\sigbl\ (left) and FWHM (right).  Distribution statistics are
also given in Table 3.

We can now compare the mass distributions from the fitted data to our
previous results (Fig. \ref{fig:SEVPdataSN}; Table 3) based on direct
measurement.  We find that low $S/N$ is somewhat mitigated by using
\sigbl\ to characterize the line width of fits to the data (Fig
\ref{fig:SEVPfitsSN}, left).  The fits allow increased precision at the
$S/N$ $\sim 10$ level, compared to measurements directly from the
spectra.  In addition, the accuracy of the VPs resulting from the fits
is also nearly unchanged down to $S/N$ $\sim 10$.  Although the fits
routinely underestimate the line peak, this does not greatly affect the
\sigbl\ results because of the insensitivity of this line
characterization to the line center.  Therefore, Gauss-Hermite fits are
advantageous for extending the usefulness of data down to $S/N$ $\sim
10$ if \sigbl\ is used to characterize the line width.

On the other hand, our fit results do not show an improvement if FWHM is
used for the line widths, at least as far as this object is concerned.
The Gauss-Hermite fits were often unable to accurately model the complex
H$\beta$ profile of NGC~5548, and the underestimation of the line peak
by the fits that was mentioned previously causes a systematic
overestimation of FWHM that increases with decreasing $S/N$.  This
overestimation of FWHM acts in the opposite direction as the trend
observed with the direct FWHM measurements from the data (i.e. a typical
underestimation of FWHM).  Therefore, as the $S/N$ decreases, a
significantly increasing difference results between the mean value of
the $M_{\rm SE}$ distributions based on direct measurement and those
based on the Gauss-Hermite fits.  From a precision standpoint, the width
of the $M_{\rm SE}$ distribution based on Gauss-Hermite fits to the
original $S/N$-level spectra is actually narrower than that of the
equivalent distribution resulting from direct measurement.  This
suggests that fitting the line profile when using FWHM may actually be
beneficial in high $S/N$ data and reduce possible systematics such as
residuals from narrow-line subtraction.  However, once the $S/N$ is
degraded, the dispersions of the distributions composed of masses
calculated from the fits (Fig. \ref{fig:SEVPfitsSN}, right) quickly
become larger than those composed of masses calculated from direct
measurement (Fig. \ref{fig:SEVPdataSN}, right).  This shows that fitting
the line profile when using FWHM does not mitigate the effects of low
$S/N$ because the fit does not accurately reproduce the true profile
shape.

For the sake of completeness, we note that there are many different
methods described in the literature for measuring FWHM that are
formulated to address issues associated with noisy data and complex line
profiles.  Here, we have chosen two methods (the formulation of
\citet{Peterson04} and the use Gauss-Hermite polynomial fits) that
differ in computational complexity and the assumptions made about the
underlying profile shape.  However, other methods also attempt to
mitigate the effects of noise.  For example, \citet{Brotherton94} define
the peak of the line based on a flux weighted mean wavelength, the
centroid, above a level that is 85\% of the line peak to decrease the
likelihood that the peak used is simply a noise spike.  Similarly,
\citet{Heckman81,Busko89} also calculate the centroid with $\gtrsim
80\%$ of the peak flux but use it in a slightly different way to
determine the line width.  Results using any of these other methods are
not expected to differ greatly from the results that we show here,
however, since our two methods effectively represent the extremes for
measuring this naively simple quantity.

\subsection{Systematic Effects Due to Blending}
\label{S_Res_decomp}

As noted earlier, the best subset of NGC~5548 spectra to use to explore
the effects of blending of spectral features is the $33$ spectra from
the Lick Observatory 3m Shane Telescope.  These are high $S/N$,
homogeneous spectra that have the broad spectral coverage necessary for
spectral decomposition.  Since spectral decomposition does not
necessarily lead to a unique solution, two independent methods were
employed as described earlier.  Cumulative distribution functions
created from the SE masses measured from the 33 Lick Observatory spectra
of NGC~5548 are shown in Figure \ref{fig:SEVPdistr_decomp}.
Distributions of $M_{\rm SE}$ are presented for all three data analysis
methods described above: the local continuum fitting method (left
panels), spectral decomposition method A (center panels), and spectral
decomposition method B (right panels).  As in previous plots, mass
results are shown for both \sigbl\ (Fig. \ref{fig:SEVPdistr_decomp}a)
and FWHM (Fig. \ref{fig:SEVPdistr_decomp}b).  Table 4 displays the
corresponding statistics for the distributions shown in Figure
\ref{fig:SEVPdistr_decomp}.

When \sigbl\ is used in the VP calculation, a full spectral
decomposition gains a small amount of precision relative to the simple,
local continuum fitting method.  More importantly though is that a
systematic offset is seen between the mean values of the local
continuum-fitted distribution versus those of decomposition methods A
and B.  Line dispersions measured from the deblended spectra (for both
methods A and B) are consistently larger than those measured using a
local continuum fit.  This is demonstrated in Figure
\ref{fig:sigbl_plot}, where we have plotted the \sigbl\ measurements
from the spectra deblended with methods A and B against those based on a
local continuum fit.  This difference is due to a combination of two
factors{\footnote{A third factor that could also lead to differing line
dispersion measurements is the presence of \feii\ emission.  Strong
\feii\ emission can obscure the line wings and line boundaries as well
as contaminate the true AGN continuum level, leading to smaller line
dispersion measurements.  Fortuitously, \feii\ emission is very weak in
NGC~5548, and therefore does not contribute to the differences observed
here.  However, this may not be the case for other objects.}.  First,
the host galaxy templates used for both decomposition methods contain a
small H$\beta$ absorption feature that effectively adds additional flux
to the center of the H$\beta$ emission line when the host is subtracted.
This absorption is not accounted for by a linear continuum fit.
However, since \sigbl\ is only weakly dependent on the line peak, this
is unlikely to make a significant contribution to the observed
difference. The second and larger contributing factor to the differences
in \sigbl\ measurements is a result of blending of
\Hbeta\ with \HeII. \citet{Decarli08} have suggested that this blending with \HeII\
complicates the measurement of the line dispersion for \Hbeta\ widths
larger that $2500$\, km s$^{-1}$.  However, we observe larger
differences for narrower \Hbeta\ widths, and therefore deduce that this
blending is a stronger function of the flux of \HeII\ rather than the
width of \Hbeta.  The effects of blending are therefore greater when the
AGN is in a higher luminosity state, when
\HeII\ is stronger, even though the \Hbeta\ line is narrower in high
states.  This is supported by the trend seen in Figure
\ref{fig:sigbl_plot} of larger \sigbl\ differences for narrower \Hbeta\
widths.

The blending of \HeII\ and \Hbeta\ could cause an overestimation of the
continuum flux level in the local continuum window defined between these
lines (see \S \ref{S_method_diff_andVPs}).  An overestimated continuum
level leads to a steeper linear fit, a subsequent over-subtraction of
the blue wing region of H$\beta$, and finally, an underestimation of
\sigbl.  The power-law continuum fit used for the decomposition
methods is not susceptible to this, since it is not fit based on local
continuum regions.  On the other hand, the \sigbl\ measurements from the
decompositions could be overestimated if some of the flux attributed to
\Hbeta\ is actually from the red wing of \HeII.  Figure
\ref{fig:profile_oplot} shows a comparison of the continuum-subtracted
mean spectrum formed from all SE spectra from each of the three data
analysis methods.  It is clear that more flux exists in the blue wing of
the deblended spectra from both methods A and B than in the spectrum
formed by subtracting the local continuum fit.  This is a consequence of
the way the continuum was fit in each case in connection with the
presence of \HeII.

Because of the large differences we observe in \sigbl\ measurements
between the decomposition methods and the local continuum fitting
method, we return to each of our decomposition methods and fit
additional contributions to account for helium emission.  Starting with
the deblended spectra we previously created with decomposition method A,
we first remove the \Hbeta\ profile by modeling the emission with a
scaled template created from a four-component Gaussian fit (two
components for the main emission and two to account for broader wings)
to \Halpha, whose blue wing is unobstructed by broad emission-line
blending. The template is fixed in velocity space and then scaled in
flux to minimize the residuals of the fit.  For these 33 spectra, the
best fits result in Balmer decrements typically in the range of
$2.8-3.2$.  The \Hbeta\ fit is then subtracted from the spectrum,
leaving the \HeII\ emission line clearly visible.  This emission is then
fit with either a single broad Gaussian profile or a double Gaussian
profile (adding a narrower component in addition to the broad component
fits 19 out of the 33 epochs better than a single component, possibly
due to residual narrow-line emission).  The best fit profile for each
epoch is subtracted from the initial, narrow-line subtracted, deblended
spectrum.  Figure \ref{fig:Hefits} (top) shows the \Hbeta\ region of the
mean spectrum formed from the SE spectra after spectral decomposition
with method A before and after subtracting the mean \HeII\ fit, which is
also shown. This method fits \HeII\ only as a means to better understand
the blending with \Hbeta.

In contrast, with method B, we return to the continuum-subtracted
spectra (i.e. after removing contributions from the host starlight,
Balmer continuum, power-law continuum, and FeII emission) and
simultaneously fit both broad and narrow optical emission lines.
Similar to the method described above for the continuum component
fitting, method B uses Levenberg-Marquardt least-squares fitting and
optimization to obtain the best overall emission-line fits to the full
spectrum.  In addition to fitting the narrow-line features as described
above, the three strongest broad Balmer lines are each fit with two
Gaussian profiles, where the best fit velocity width is held fixed for
all three lines.  Both \heii\ and \hei\ emission lines are fit with a
single Gaussian profile, and although these widths are not tied to the
Balmer line widths, the widths of \HeII\ and the \hei\ emission under
the \Hbeta, \ob\ region are tied to the width of the unblended \HeI\
line in the same way the Balmer line widths are tied together.  Each set
of emission lines of a given species and type of emission (i.e. narrow
or broad) is isolated in the total fit, so that only the emission of
interest can be subtracted.  Since the narrow-line emission was
subtracted previously, we now subtract the broad helium emission,
effectively deblending \Hbeta\ from \HeII.  Figure \ref{fig:Hefits}
(bottom) shows the \Hbeta\ region of the mean spectrum created from the
33 SE spectra after decomposition with method B before and after
subtracting the average helium fit, which is also shown.

We measure line widths in these He-deblended spectra with a
newly-defined blue boundary for \Hbeta\ at $4720$\AA\ (compared to
$4747$\AA\ previously).  This boundary was extended because the edge of
the blue wing of \Hbeta\ is better discerned without the presence of
\HeII.  Figure \ref{fig:noHe_sigbl_plot} shows new \sigbl\ measurements
for the He-deblended \Hbeta\ line from the two decomposition methods
compared again to \sigbl\ from the local continuum method.  The \sigbl\
measurements from method A still disagree with the local continuum
fitting method as much as, if not more than, before subtraction of
\HeII.  However, the new \sigbl\ measurements from method B are now
consistent with the local continuum fitting method.

The observed differences in these new \sigbl\ measurements between
method A and method B come from the procedure and assumptions that each
method uses to fit the spectral emission lines.  The line widths from
method B now agree with the local continuum fitting method because the
combined best fit to both lines tends to result in an \Hbeta\ profile
that basically sits on top of a broad \HeII\ profile.  In the wavelength
region between the two emission lines (i.e. where the local continuum is
defined), the difference between the continuum level and the flux level
observed in the blended spectrum is usually attributed completely to
\HeII\ emission by method B.  Therefore, when \HeII\ is subtracted, the
flux level of this region is reduced nearly to the level of the
continuum, which is what is assumed by the local continuum fitting
method, thus making these two methods consistent.  On the other hand,
method A subtracts the \Hbeta\ with an \Halpha\ template before fitting
\HeII.  Because the \Halpha\ profile has very extended wings, this method
necessarily assumes that \Hbeta\ also has this extended, broad component.
Therefore, nearly opposite to method B, method A effectively fits a
\HeII\ profile that is sitting on top of a very broad \Hbeta\
profile and consequently subtracts a smaller \HeII\ component.  This
results in \sigbl\ measurements that are equally or even more inconsistent
with previous measurements because it extends the \Hbeta\ wing under the
\HeII\ profile.  Because this extended blue wing is hidden under \HeII,
method A results suggest that the local continuum fitting method is
significantly underestimating \sigbl\ (by as much as $40 \%$).

Evidence suggests that the helium lines are consistently broader than
the Balmer lines in Type 1 AGNs \citep{Osterbrock82}.  This is always
the case in the rms spectrum of AGNs that have been monitored for
reverberation mapping studies, as well.  Additionally, in the few cases
for which reverberation lags could be measured for \HeII\, the lags are
shorter than the corresponding \Hbeta\ lag in the same object
\citep[see][]{Peterson04}.  This suggests that given the virial
hypothesis for a single source, the material responsible for \HeII\
emission is closer to the central source than that responsible for the
\Hbeta\ emission and moving at a faster velocity, thus producing broader
emission lines.  Method B supports this evidence with the emission line
models and results described above.  On the other hand, although the
fits of method A do not reproduce the same broad \HeII\ emission, the
assumption this method makes about the similarities that should exist
between the shape of the \Halpha\ and \Hbeta\ profiles are hard to
discount, given that these two species should exist in similar regions
of the BLR.  Instead, our analysis demonstrates that there is not a
unique method to account for the blending of \Hbeta\ and \HeII\ that
results in consistent line dispersion measurements of \Hbeta.
Therefore, we conclude that this blending is a potential problem for the
use of \sigbl\ in calculating $M_{\rm SE}$.

Blending is less likely to be a limitation for reverberation mapping
studies that use the line dispersion measured in the rms spectrum,
however.  Blending between \Hbeta\ and \HeII\ is often lessened in the
rms spectrum because the broad wings of the lines that are the most
blended tend not to be as variable as the more central parts of the
line.  To test this, we characterized the \Hbeta\ line width with
\sigbl\ in the 3 rms spectra formed from the three sets of spectra
created during the deblending analysis (after the local continuum fit,
decomposition method A, and decomposition method B).  We did not account
for \HeII\ emission in the rms spectrum before measuring \sigbl\ in the
local continuum subtracted rms spectrum.  However, the \HeII\ emission
in the two rms spectra formed after decomposition methods A and B was
modeled with a single Gaussian profile and subtracted.  We find that
measurements of \sigbl\ from the rms spectra from all three methods are
consistent to within $1\sigma$.  This consistency suggests that the
masses determined through reverberation studies that use the line
dispersion measured from the rms spectrum are not as susceptible as SE
masses to this bias in \sigbl\ caused by blending.  Additionally, it is
worth noting that not all AGNs have strong blending of \HeII\ and
\Hbeta, superceding the need for such caution with the use of \sigbl.

Different concerns arise when FWHM is used to characterize the \Hbeta\
line width.  Figure \ref{fig:SEVPdistr_decomp}b demonstrates that all
three methods are in agreement, on average, with equally good precision
and moderately small offsets from their respective reverberation results
(given in Table 4).  The small systematic difference between the mean SE
masses of the decomposition methods and the local continuum fit is most
likely due to the small \Hbeta\ absorption feature present in the host
galaxy light, as discussed above.  Figure \ref{fig:fwhm_plot} shows that
FWHM, unlike the line dispersion, is less sensitive to the details of
measurement, however.  The differences seen in the line dispersion
measurements are not present for the FWHM measurements, since blending
in the wings and the definition of the continuum have a much smaller
effect on the FWHM value.  However, these general observations and the
FWHM statistics in Table 4 exclude the outliers at the low-mass end of
the distributions in Figure \ref{fig:SEVPdistr_decomp}b, shown by the
thin black curves (also labeled in Fig. \ref{fig:fwhm_plot}).  These
points are outliers because of a particularly complex line profile,
characterized by an asymmetric red bump, present in these two epochs
(JD2452030 and JD2452045).  These epochs illustrate that FWHM can be
complicated by profile features such as the gross asymmetries and double
peaks that the broad Balmer lines sometimes exhibit.

Figure \ref{fig:outlier_oplot} shows how FWHM is defined for the \Hbeta\
profile on JD2452030 for each of the two decomposition methods and for
a local continuum fit.  In each case, we measure FWHM following the
procedure of \citet{Peterson04}.  The differences in the FWHM
measurements for this spectrum are due in part to the complex profile of
this line and in part to the differences in the peak flux of the line
for the different methods.  Figure \ref{fig:outlier_oplot} shows that
each of the three methods removes slightly different amounts of
narrow-line emission.  These small differences change the total flux in
the line by at most a few percent, but the change in the line peak
combined with the complex profile are sufficient to cause large
differences in the measurements of FWHM, and thus $M_{\rm SE}$.

Despite the observed differences in the SE \sigbl\ measurements between
each decomposition method after accounting for \HeII\ blending, masses
derived from both methods otherwise differ very little.  The dispersions
in the SE mass distributions from both methods are nearly equal, however
masses derived with the use of method A seem somewhat more accurate,
with smaller $\langle \Delta {\rm log} M\rangle$ values than method B.

\section{Discussion and Conclusion}
\label{S_Discuss_and_conclude}

We have undertaken a careful examination of some of the systematics
associated with measurements of emission-line widths for the purpose of
calculating black hole virial masses from single-epoch spectra.  The
systematics on which we focused our attention are (i) intrinsic AGN
variability, (ii) contributions by constant spectral components, (iii)
$S/N$ of the data, and (iv) blending with the different spectral
components, particularly the underlying host galaxy.

Throughout this analysis we have not displayed a preference for either
the line dispersion or the FWHM to characterize the line width and have
instead shown that there are both advantages and limitations to each
measure.  Specifically, FWHM provides consistent results for lower $S/N$
spectra without the use of profile fits, and it is much more robust in
the presence of blending.  However, FWHM should only be used in spectra
that have had the narrow line components carefully removed, as the
sensitivity of FWHM to the presence and/or removal method of narrow
emission lines is a serious limitation.  On the other hand, the line
dispersion is advantageous in this respect, since it is rather
insensitive to the details of narrow-line component subtraction.
However, its use should be limited to data characterized by relatively
high $S/N$ or with profile fits to the emission lines.  Unlike FWHM, the
greatest limitation of using the line dispersion is blending in the line
wings, and use of the line dispersion should therefore be avoided if
there is emission line blending that has not been modeled and removed.
As we have shown here, however, even in the case of modeling, the
accuracy of the model may be questionable.  In the case of NGC~5548, if
decomposition model B is correct (i.e., where the \HeII\ line is fit
assuming the same velocity width as the unblended \HeI\ line), then
correcting for the blending of \Hbeta\ and \HeII\ by modeling and
subtracting the helium emission produces consistent results with the
local continuum-fitting method.  However, if method A is the more
accurate representation of the blending (i.e., where the \HeII\ line was
modeled assuming the line profile of \Hbeta\ is the same as \Halpha),
then there will be a resulting mean offset in the SE masses of $\sim
0.1$ dex compared to the local continuum-fitting method due to
underestimation of the blended \Hbeta\ line dispersion in the latter
method.  Because of these difficulties, when blending complicates the
line profile shape or boundaries of SE spectra, it is best to use FWHM.

To summarize the effects of these systematics on SE masses, Table 5
gives an error budget displaying how each systematic affects the
uncertainties in SE mass estimates in terms of increasing or decreasing
the precision and accuracy of the measurement, where we generalize our
results here to both low luminosity Seyfert-type AGNs and quasars.
While nearly all of the systematic uncertainties we investigated add to
the dispersion in the SE mass distributions in varying amounts, some
effects also cause often severe systematic shifts in the distributions,
leading to overall under- or overestimations of SE masses.  Readers
should be particularly cautious about these effects because large
statistical studies cannot average out these types of systematics.  In
summarizing the sources of error covered here, we use the same
description of the precision and accuracy as above, with the accuracy
described as an offset in the mean SE virial mass, and the precision
described by the dispersion in the mass distribution.  In Table 5,
however, we assume that errors are independent and the distributions are
close enough to Gaussian that we can add independent errors in
quadrature to determine the cumulative effect.  We therefore describe
the additional offset and dispersion due to each systematic with respect
to the SE mass calculation which results in the minimum observed
uncertainties (i.e., Fig. \ref{fig:SEVPdistr_var}).  In Table 5 we
consider the following individual sources of error in the SE masses for
both characterizations of the line width:

\begin{enumerate}

\item Random measurement errors.  These are simply due to inherent
uncertainties in any measurement of luminosity and line width.
Empirically, we determine these uncertainties by comparing measurements
of closely spaced observations, assuming that these parameters change
little over very short time scales (i.e., time scales much shorter than
the reverberation time scale).  We use this empirical method to estimate
the uncertainties for the line width and luminosity of the NGC~5548 data
set, which we propagate through to determine uncertainties in the mass
estimates, listed in Table 5.  Uncertainties are not listed for quasars
because the size of the PG1229 data set is much smaller and with fewer
closely spaced observations than that of NGC~5548.  We could therefore
not accurately estimate uncertainties in this manner.  However, given
the small observed dispersion in the SE virial masses for PG1229
($\sim0.05$ dex), measurement uncertainties are likely to be very small.

\item Variability on reverberation timescales (see
Fig. \ref{fig:deltalogM} and Fig. \ref{fig:SEVPdistr_var}b).  Our
analysis on re\-ver\-ber\-a\-tion-timescale variability shows that SE
spectra can reproduce the reverberation-based virial product that would
be measured at the same time to about 0.10 dex (i.e., $\sim25$\%) for
Seyferts and to about 0.05 dex (i.e., $\sim15$\%) for quasars.  This is
an interesting result, given the quadrature sum of the individual
dispersions in luminosity and line width for NGC~5548 add to be $\sim
0.17$ dex, regardless of line width measure.  This is significantly
larger than the dispersion in the virial masses, and therefore confirms
the presence of a virial relation between the line width and luminosity
(i.e., the BLR radius).  An additional ramification for quasars is that
the dispersion for PG1229 determined here represents more than a factor
of two less than even the formal, observational uncertainties in the
reverberation mass for this object.  This suggests that once the zero
point and slope of scaling relations such as the $R-L$ relation are
accurately determined, it may be more accurate to simply use the scaling
relations to determine masses of individual sources than to make direct
mass measurements.  It also follows that SE mass estimates can then
easily be acquired with relative certainty for high redshift objects, as
long as the extrapolation of the scaling relations to these luminosity
regimes is valid.

\item Longer-term secular variations.  At least in the case of NGC~5548,
we see that longer-term (dynamical timescale?) variations cause changes
in both the SE and re\-ver\-ber\-a\-tion-based virial product. The
amplitude of these variations is similar to those on reverberation time
scales, creating an additional dispersion in the SE virial products of
about $0.09$ or $0.05$ dex for FWHM and \sigbl, respectively.  This
longer-term secular variability adds to the reverberation-scale
variability described above (and seen in Figure \ref{fig:deltalogM}) to
produce the observed dispersion (Fig. \ref{fig:SEVPdistr_var}a) in the
SE virial product for Seyferts. We cannot estimate the contribution of
secular variations to the observed dispersion for quasars, since all the
data available for PG1229 was used in a single reverberation experiment
and observations did not span dynamical timescales for this object.

\item Combined minimum uncertainty.  The combination of the above effects
sets a ``minimum observable uncertainty'' for SE-based masses, given in
line 4 of Table 5 (see also Fig. \ref{fig:SEVPdistr_var}).  Adding
measurement uncertainties, reverberation timescale variability, and
longer-term secular variability effects in quadrature yields an estimate
of the observable dispersion in SE masses for Seyferts of $0.12-0.16$
dex ($\sim30-45$\%) and less for quasars (although the long-term secular
effects are unexplored in this case).

\item Failure to remove host galaxy starlight.  Host galaxy contamination
causes an overestimation of the luminosity and thus the mass.  The
effect of this contamination on the precision of SE mass estimates is
minimal so it does not further broaden the distribution of mass
measurements.  Instead, it affects the accuracy of the mass estimate,
resulting in an additional mean offset, listed in Table 5, compared to
the offset observed when the host contamination is removed (Also compare
the mean distribution values of Figs. \ref{fig:SEVPdistr_var} and
\ref{fig:SEVPnl_nogs}).  The size of the systematic overestimation of
the mass depends on the fraction of host starlight contamination,
however.  Since both NGC~5548 and PG1229 lie near the middle of the
sample of AGNs used to set the slope of the $R-L$ relation, the effect
due to host-galaxy contamination could be much worse or more minimal
depending on whether the luminosity is much smaller or larger
(respectively) than the objects presented here
\citep{Bentz08}.

\item Failure to remove narrow \Hbeta.  The \Hbeta\ narrow emission-line
component is by far the biggest source of error for both Seyferts and
quasars when using FWHM to characterize the line width, adding
significantly to the dispersion and offset, as shown in Table 5.  In
particular, this offset causes SE masses to be underestimated by nearly
an order of magnitude for Seyfert-type galaxies that often have strong
narrow-line components.  Notice, however, that because of the
insensitivity of \sigbl\ to the line center, this effect increases the
dispersion of the SE mass distributions very little or not at all when
the line width is characterized by \sigbl.  However, the systematic
offset for the \sigbl\ case is still nearly doubled compared to the
offset observed for the minimum uncertainty case.  This makes it
imperative to remove narrow line components before measuring line
widths, regardless of how the line width is characterized.

\item Limitations due to $S/N$.  Direct measurements from low $S/N$
spectra add an additional systematic offset in the SE mass measurements
because of a systematic underestimation of the line width, as well as
decreased precision in these measurements.  Our fits to the line
profiles do increase the usefulness of $S/N$-level $\sim10$ spectra with
\sigbl.  However, they generally make things worse for FWHM, leading to lower
precision masses than when direct measurements of the line widths are
used, as well as systematic overestimations of the line width and mass,
an effect that is opposite to that observed when measuring FWHM directly
from the data.  Therefore, to avoid either underestimating SE masses
when measuring line widths directly from the data or overestimating SE
masses when line profiles are fit, SE mass studies should be conducted
using high $S/N$ ($\gtrsim 20$ pixel$^{-1}$) spectra.

\end{enumerate}

\acknowledgements
The authors are grateful to C. A. Onken for the use of his Gauss-Hermite
fitting software.  We would also like to thank David Weinberg for many
useful comments and suggestions which improved several aspects of this
paper.  This work has been supported by the NSF through grant
AST-0604066 and by NASA through grant AR-10691 from the Space Telescope
Science Institute, which is operated by AURA, Inc., under NASA contract
NAS5-26555.  This research has made use of the NASA/IPAC Extragalactic
Database (NED) which is operated by the Jet Propulsion Laboratory,
California Institute of Technology, under contract with the National
Aeronautics and Space Administration.




\clearpage


\begin{deluxetable}{lccccccc}
\label{T_variability_stats}
\rotate
\tablecolumns{8}
\tablewidth{0pt}
\tablecaption{Systematic Effects due to Variability}
\tablehead{
\multicolumn{2}{c}{}&
\multicolumn{3}{c}{$M$ ($M_{\odot}$) $\propto$ \sigbl}&
\multicolumn{3}{c}{$M$ ($M_{\odot}$) $\propto$ FWHM}\\
\cline{3-8}
\colhead{Object}&
\colhead{$N_{\rm SE}$}&
\colhead{log\,$M_{\rm vir}$}&
\colhead{$\langle \rm{log}\,M_{\rm SE} \rangle \pm \sigma_{\rm SE}$}&
\colhead{$\langle \Delta {\rm log} M \rangle$}&
\colhead{log\,$M_{\rm vir}$}&
\colhead{$\langle \rm{log}\,M_{\rm SE} \rangle \pm \sigma_{\rm SE}$}&
\colhead{$\langle \Delta {\rm log} M \rangle$}
}

\startdata

NGC~5548   &370& 7.21$\pm$0.02 & $7.12\pm 0.12$ & $-0.09$ & $8.06 \pm 0.02$ & $7.95 \pm 0.16$ & $-0.11$ \\
PG1229+204 & 33& 7.28$\pm$0.25 & $7.22\pm 0.05$ & $-0.06$ & $8.03 \pm 0.25$ & $7.92 \pm 0.06$ & $-0.11$ \\

\enddata

\end{deluxetable}

\clearpage

\begin{deluxetable}{lccccccccc}
\label{T_const_compon_stats}
\rotate
\tablecolumns{10}
\tablewidth{0pt}
\tablecaption{Systematic Effects due to Constant Components}
\tablehead{

\multicolumn{4}{c}{}&
\multicolumn{3}{c}{$M$ ($M_{\odot}$) $\propto$ \sigbl}&
\multicolumn{3}{c}{$M$ ($M_{\odot}$) $\propto$ FWHM}\\
\cline{5-7}
\cline{8-10}
\multicolumn{2}{c}{}&
\colhead{Narrow}&
\colhead{Host}&
\colhead{}&
\colhead{$\langle {\rm log}\,M_{\rm SE} \rangle$}&
\multicolumn{2}{c}{}&
\colhead{$\langle {\rm log}\,M_{\rm SE} \rangle$}&
\colhead{}\\
\colhead{Object}&
\colhead{$N_{\rm SE}$}&
\colhead{Lines}&
\colhead{Starlight}&
\colhead{log\,$M_{\rm vir}$}&
\colhead{$\pm \sigma_{\rm SE}$}&
\colhead{$\langle \Delta {\rm log}\,M \rangle$}&
\colhead{log\,$M_{\rm vir}$}&
\colhead{$\pm \sigma_{\rm SE}$}&
\colhead{$\langle \Delta {\rm log}\,M \rangle$}
}

\startdata

NGC~5548\tablenotemark{a}  &370&present &present &7.21$\pm$0.02& $7.21\pm 0.11$ & $+0.00$ & $8.06\pm 0.02$ & $7.22\pm 0.47$ & $-0.84$ \\
NGC~5548\tablenotemark{b}  &370&removed &present &7.21$\pm$0.02& $7.29\pm 0.11$ & $+0.08$ & $8.06\pm 0.02$ & $8.12\pm 0.14$ & $+0.06$ \\
NGC~5548\tablenotemark{c}  &370&present &removed &7.21$\pm$0.02& $7.05\pm 0.13$ & $-0.16$ & $8.06\pm 0.02$ & $7.06\pm 0.52$ & $-1.00$ \\
PG1229...\tablenotemark{a}& 33&present &present &7.28$\pm$0.25& $7.33\pm 0.05$ & $+0.05$ & $8.03\pm 0.25$ & $8.00\pm 0.07$ & $-0.03$ \\
PG1229...\tablenotemark{b}& 33&removed &present &7.28$\pm$0.25& $7.34\pm 0.05$ & $+0.06$ & $8.03\pm 0.25$ & $8.04\pm 0.06$ & $+0.01$ \\
PG1229...\tablenotemark{c}& 33&present &removed &7.28$\pm$0.25& $7.21\pm 0.05$ & $-0.07$ & $8.03\pm 0.25$ & $7.88\pm 0.07$ & $-0.15$ \\

\enddata

\tablenotetext{a}{Refer to Figure \ref{fig:SEVP_nocorrections}.}
\tablenotetext{b}{Refer to Figure \ref{fig:SEVPnl_nogs}.}
\tablenotetext{c}{Refer to Figure \ref{fig:SEVPdistr_wlgs}.}

\tablecomments{See Table 1 for the case in which both the narrow emission
lines and the host starlight are removed for the virial mass
calculations for both NGC~5548 and PG1229.}

\end{deluxetable}

\clearpage

\begin{deluxetable}{ccccccccc}
\label{T_SN_stats}
\rotate
\tablecolumns{9}
\tablewidth{0pt}
\tablecaption{Systematic Effects due to Signal-to-Noise Ratio}
\tablehead{

\multicolumn{3}{c}{}&
\multicolumn{3}{c}{$M$ ($M_{\odot}$) $\propto$ \sigbl}&
\multicolumn{3}{c}{$M$ ($M_{\odot}$) $\propto$ FWHM}\\
\cline{4-9}

\colhead{Data or}&
\multicolumn{3}{c}{}&
\colhead{$\langle {\rm log}\,M_{\rm SE} \rangle$}&
\multicolumn{2}{c}{}&
\colhead{$\langle {\rm log}\,M_{\rm SE} \rangle$}&
\colhead{}\\

\colhead{Fit}& 
\colhead{$S/N$}&
\colhead{$N_{\rm SE}$}&
\colhead{log\,$M_{\rm vir}$}&
\colhead{$\pm \sigma_{\rm SE}$}&
\colhead{$\langle \Delta {\rm log}\,M \rangle$}&
\colhead{log\,$M_{\rm vir}$}&
\colhead{$\pm \sigma_{\rm SE}$}&
\colhead{$\langle \Delta {\rm log}\,M \rangle$}
}

\startdata

Data & Orig  &$270$&$7.23\pm0.02$&$7.12\pm0.14$&$-0.11$&$8.02\pm0.02$&$7.96\pm0.19$&$-0.06$\\
Data & $\sim 20$&$270$&$7.23\pm0.02$&$7.11\pm0.15$&$-0.12$&$8.02\pm0.02$&$7.93\pm0.18$&$-0.09$\\
Data & $\sim 10$&$270$&$7.23\pm0.02$&$7.09\pm0.22$&$-0.14$&$8.02\pm0.02$&$7.85\pm0.19$&$-0.17$\\
Data & $\sim 05$&$270$&$7.23\pm0.02$&$7.04\pm0.31$&$-0.19$&$8.02\pm0.02$&$7.84\pm0.21$&$-0.18$\\
Fit & Orig  &$270$&$7.23\pm0.02$&$7.11\pm0.14$&$-0.12$&$8.02\pm0.02$&$8.03\pm0.17$&$+0.01$\\
Fit & $\sim 20$&$270$&$7.23\pm0.02$&$7.10\pm0.17$&$-0.13$&$8.02\pm0.02$&$8.08\pm0.23$&$+0.06$\\
Fit & $\sim 10$&$270$&$7.23\pm0.02$&$7.10\pm0.17$&$-0.13$&$8.02\pm0.02$&$8.08\pm0.23$&$+0.06$\\
Fit & $\sim 05$&$270$&$7.23\pm0.02$&$7.06\pm0.28$&$-0.17$&$8.02\pm0.02$&$8.13\pm0.29$&$+0.11$\\
	     
\enddata     
	     
\end{deluxetable}

\clearpage

\begin{deluxetable}{lcccccccc}
\label{T_decomp_stats}
\rotate
\tablecolumns{9}
\tablewidth{0pt}
\tablecaption{Systematic Effects due to Blending}
\tablehead{

\multicolumn{3}{c}{}&
\multicolumn{3}{c}{$M$ ($M_{\odot}$) $\propto$ \sigbl}&
\multicolumn{3}{c}{$M$ ($M_{\odot}$) $\propto$ FWHM}\\
\cline{4-9}

\multicolumn{2}{c}{}&
\colhead{Decomposition}&
\colhead{}&
\colhead{$\langle {\rm log}\,M_{\rm SE} \rangle$}&
\multicolumn{2}{c}{}&
\colhead{$\langle {\rm log}\,M_{\rm SE} \rangle$}&
\colhead{}\\

\colhead{Object}&
\colhead{$N_{\rm SE}$}&
\colhead{Method}&
\colhead{log\,$M_{\rm vir}$}&
\colhead{$\pm \sigma_{\rm SE}$}&
\colhead{$\langle \Delta {\rm log}\,M \rangle$}&
\colhead{log\,$M_{\rm vir}$}&
\colhead{$\pm \sigma_{\rm SE}$}&
\colhead{$\langle \Delta {\rm log}\,M \rangle$}
}

\startdata

NGC~5548&33&Local Cont. Fit&$7.22\pm 0.02$&$7.23\pm 0.13$&$+0.01$&$8.16\pm 0.02$&$8.16\pm 0.09$&$+0.00$ \\
NGC~5548&33&Method A&$7.32\pm 0.02$&$7.31\pm 0.10$&$-0.01$&$8.11\pm 0.02$&$8.09\pm 0.08$&$-0.02$ \\
NGC~5548&33&Method B&$7.38\pm 0.02$&$7.31\pm 0.09$&$-0.07$&$8.15\pm 0.02$&$8.08\pm 0.09$&$-0.07$ \\

\enddata

\end{deluxetable}

\clearpage

\begin{deluxetable}{lrcrcrcrc}
\label{T_summary}
\rotate
\tablecolumns{9}
\tablewidth{0pt}
\tablecaption{Individual Error Sources for SE Mass Measurements}
\tablehead{

\colhead{}&
\multicolumn{4}{c}{Seyfert}&
\multicolumn{4}{c}{Quasar}\\
\cline{2-9}

\colhead{}&
\multicolumn{2}{c}{$M$ $\propto$ FWHM}&
\multicolumn{2}{c}{$M$ $\propto$ \sigbl}&
\multicolumn{2}{c}{$M$ $\propto$ FWHM}&
\multicolumn{2}{c}{$M$ $\propto$ \sigbl}\\

\colhead{Effect on $M_{\rm SE}$}&
\colhead{offset}&
\colhead{dispersion}&
\colhead{offset}&
\colhead{dispersion}&
\colhead{offset}&
\colhead{dispersion}&
\colhead{offset}&
\colhead{dispersion}
}

\startdata
Random measurement error:                         &\nodata&$0.07$&\nodata&$0.04$&\nodata&\nodata\tablenotemark{a}&\nodata&\nodata\tablenotemark{a}\\
Variability (RM timescales):                      &$-0.11$&$0.11$&$-0.10$&$0.10$&$-0.11$&$0.06$&$-0.06$&$0.05$\\
Longer term secular variations +\\
\hspace{8pt}slight inhomogeneity of spectra:      &$-0.00$&$0.09$&$+0.01$&$0.05$&\nodata\tablenotemark{b}&\nodata\tablenotemark{b}&\nodata\tablenotemark{b}&\nodata\tablenotemark{b}\\
Above effects (min. uncertainty):                 &$-0.11$&$0.16$&$-0.09$&$0.12$&$-0.11$&$0.06$&$-0.06$&$0.05$\\
Additional systematics:\\
\hspace{4pt}Failure to remove host galaxy:        &$+0.17$&$-0.08$&$+0.17$&$-0.05$&$+0.12$&$0.00$&$+0.12$&$0.00$\\
\hspace{4pt}Failure to remove narrow \Hbeta:      &$-0.89$&$0.49$&$-0.07$&$0.05$&$-0.04$&$0.04$&$-0.01$&$0.00$\\
\hspace{4pt}$S/N$ limitation (data, $S/N$=10):    &$-0.06$&$0.10$&$-0.05$&$0.18$&\nodata&\nodata&\nodata&\nodata\\
\hspace{86pt}(data, $S/N$=05):                    &$-0.07$&$0.14$&$-0.10$&$0.29$&\nodata&\nodata&\nodata&\nodata\\
\hspace{4pt}$S/N$ limitation (fit, $S/N$=20):     &$+0.17$&$0.17$&$-0.04$&$0.12$&\nodata&\nodata&\nodata&\nodata\\
\hspace{86pt}(fit, $S/N$=10):                     &$+0.17$&$0.17$&$-0.04$&$0.12$&\nodata&\nodata&\nodata&\nodata\\
\hspace{86pt}(fit, $S/N$=05):                     &$+0.22$&$0.24$&$-0.08$&$0.25$&\nodata&\nodata&\nodata&\nodata\\

\enddata

\tablenotetext{a}{Uncertainties could not be determined; see Discussion,
\S\ref{S_Discuss_and_conclude}, individual sources of error (1).}
\tablenotetext{b}{Uncertainties could not be determined; see Discussion,
\S\ref{S_Discuss_and_conclude}, individual sources of error (3).}

\end{deluxetable}
\clearpage

\begin{figure}
\epsscale{1}
\plotone{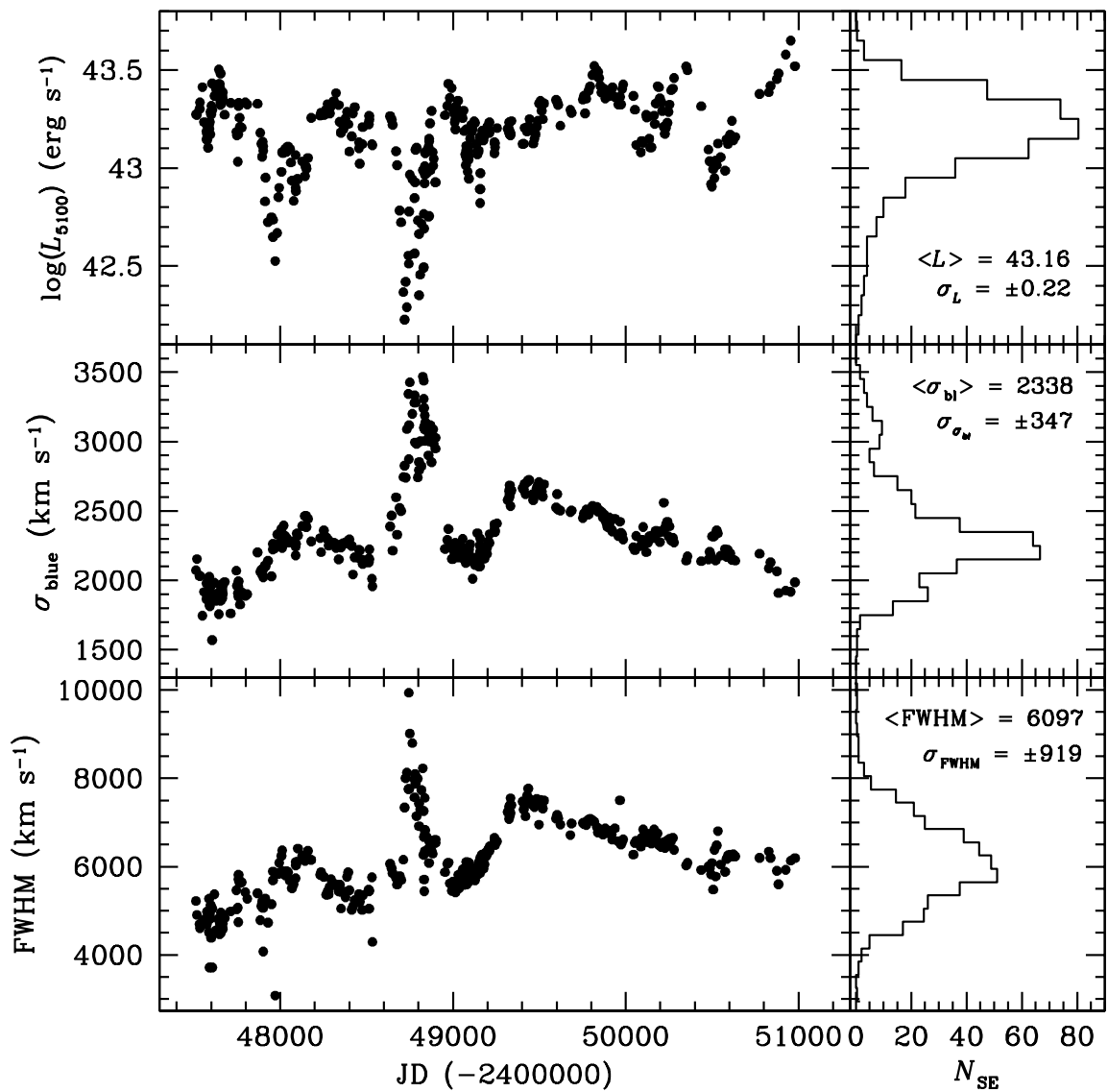}
\caption{Starlight-corrected luminosity and narrow-line subtracted \Hbeta\
line width measurements from the full set of NGC~5548 spectra.  Left
panels show individual SE measurements as a function of time, and right
panels show distributions of each measured quantity, with the mean and
standard deviation of the sample given.}

\label{fig:VandLvsTime}
\end{figure}
\clearpage

\begin{figure}
\epsscale{.95}
\plotone{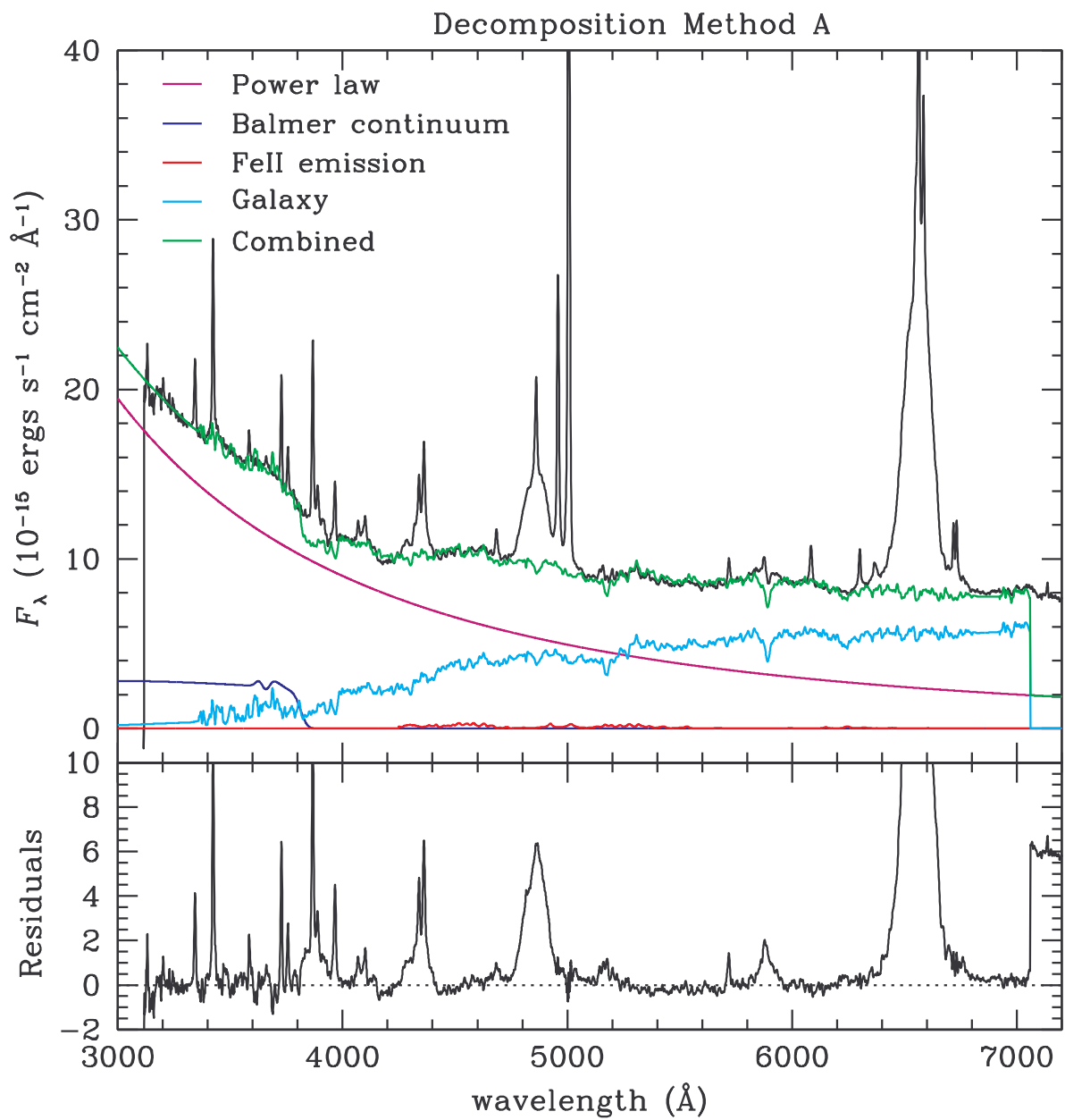}
\caption{The multi-component fit to a typical spectrum of NGC~5548 for
decomposition method A. In the top panel, the rest-frame spectrum shown
together with a four-component fit: a power-law continuum, a host-galaxy
spectrum, Balmer continuum emission, and weak optical \feii\
emission. The combined fit is displayed in green.  In the bottom panel,
the corresponding residual spectrum is presented after additional
subtraction of the narrow emission-line components (not shown).}

\label{fig:MDdecompFig}
\end{figure}
\clearpage

\begin{figure}
\epsscale{.95}
\plotone{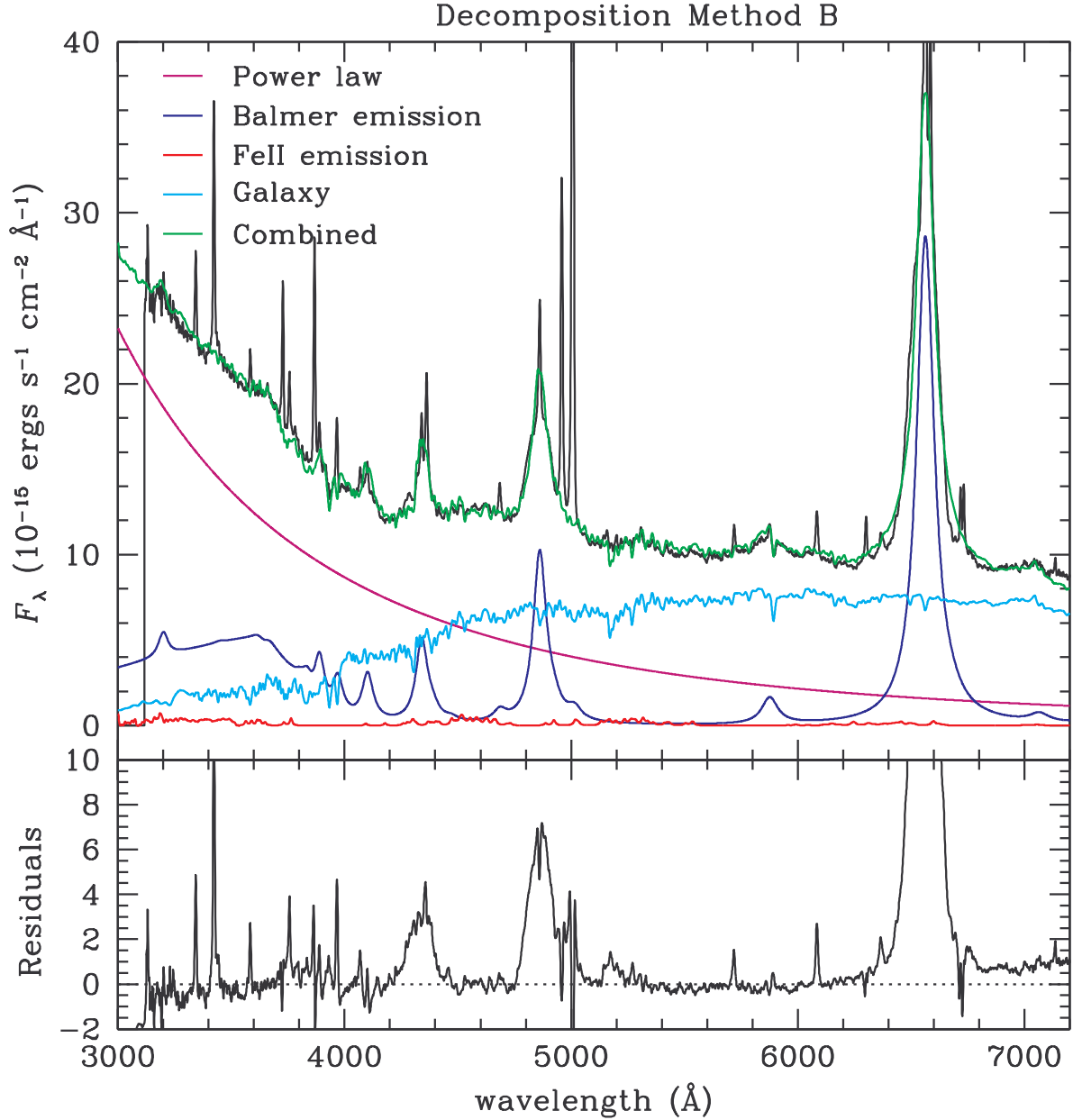}

\caption{The multi-component fit to a typical spectrum of NGC~5548 for
decomposition method B. In the top panel the rest frame spectrum shown
has been corrected for Galactic extinction and is shown together with a
four-component fit: a power-law continuum, a host-galaxy spectrum,
Balmer continuum and broad-line emission, and weak optical \feii\
emission. The combined fit is displayed in green.  Note: the Balmer line
emission shown here is only included to prevent the fitting routine from
attempting to assign continuum emission components to the profile wings
and is not included in the final fit that is subtracted to create the
residual spectrum (bottom).  In addition, the residual spectrum
presented also includes additional subtraction of narrow emission-line
components, not shown.}

\label{fig:MVdecompFig}
\end{figure}
\clearpage

\begin{figure}
\epsscale{.94}
\plotone{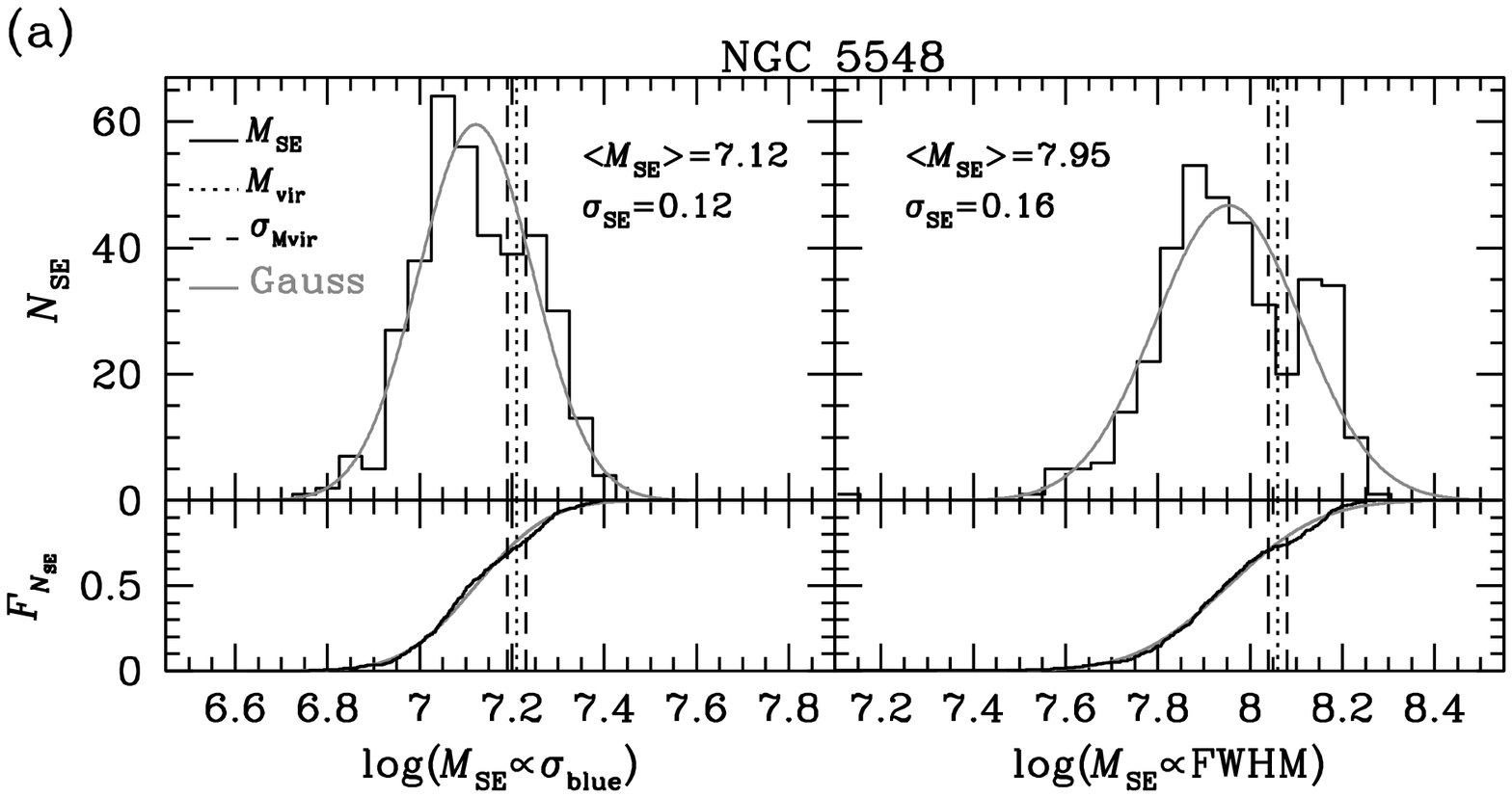}
\plotone{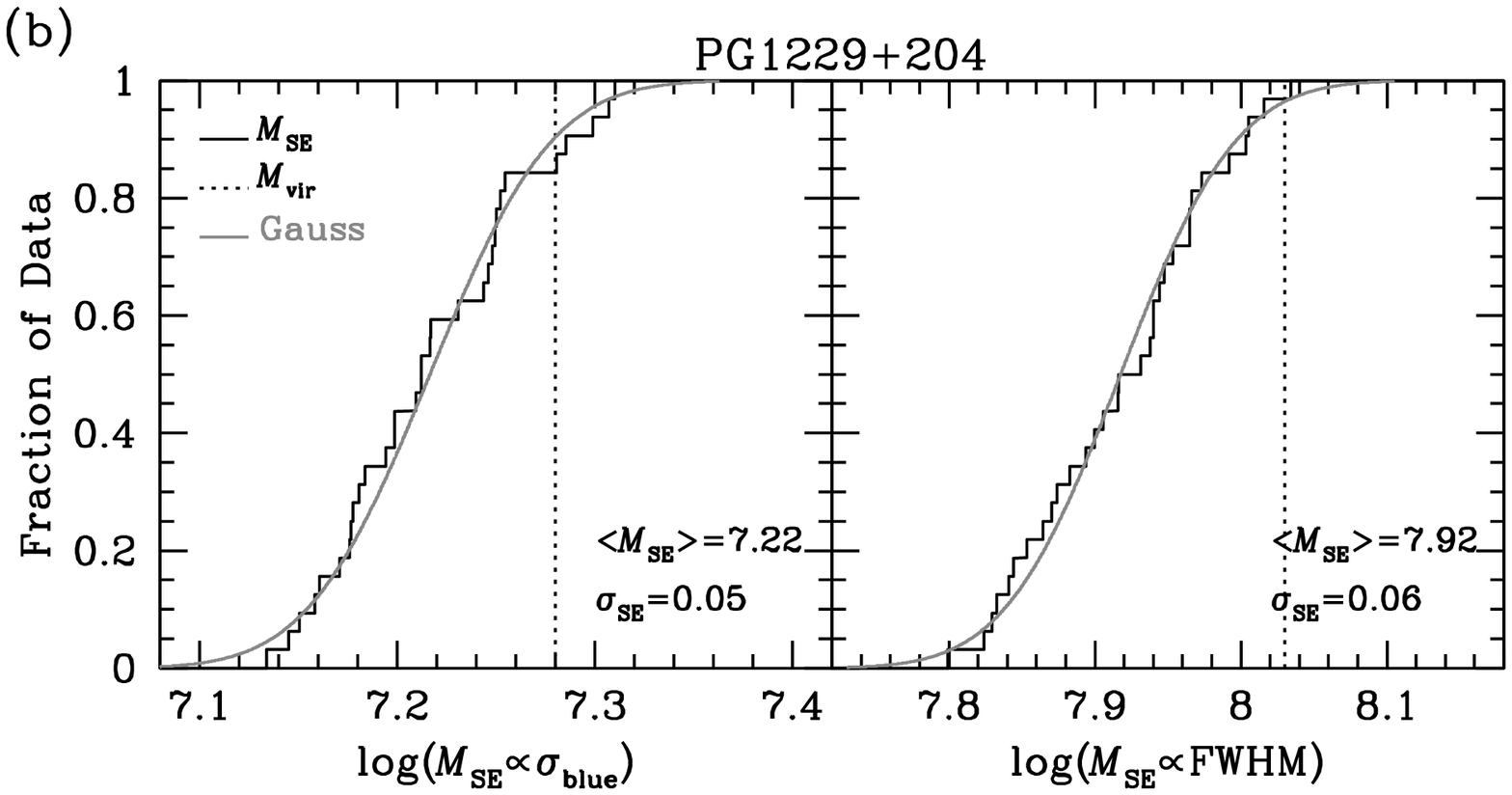}

\caption{Virial mass distributions for the full NGC~5548 (a) and
PG1229 (b) data sets.  The Solid lines show the distributions of virial
masses calculated with equation 2 using both \sigbl\ (left) and the FWHM
(right) to measure \Hbeta\ line widths: histograms for the larger
NGC~5548 data set and cumulative distribution functions (CDFs) for the
smaller PG1229 data set.  Narrow lines and host galaxy starlight have
been subtracted from all spectra before calculating masses.  A Gaussian
function with the same mean, dispersion, and area as the data is
overplotted in gray.  The distribution mean and dispersion is shown in
each plot, where values listed have not been corrected for random
measurement uncertainties (see \S \ref{S_Res_variability}).  For each data
set and line width measure, the vertical lines represent the
reverberation virial mass (dotted) with measurement uncertainties
(dashed; not shown for PG1229 because they are typically larger than the
widths of the distributions).}

\label{fig:SEVPdistr_var}
\end{figure}

\clearpage

\begin{figure}
\epsscale{1}
\plotone{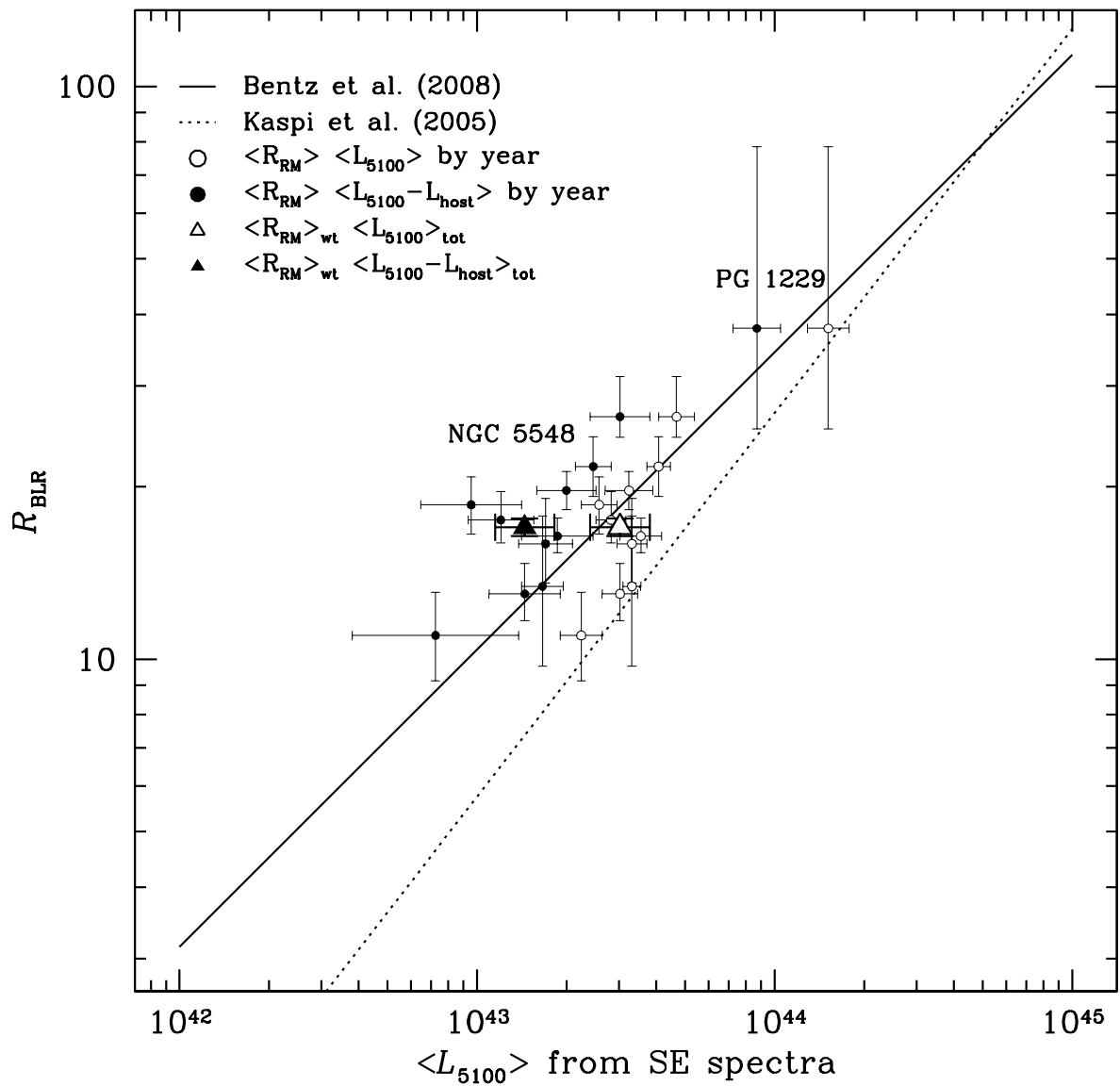}

\caption{Broad line region radius-luminosity relationship for the PG1229 data
and weighted mean as well as individual years of NGC~5548 data.  Points
are plotted for luminosities both before and after subtracting the host
galaxy starlight contribution to the $5100$ \AA\ continuum flux. The
\citet{Bentz08} relation and the \citet{Kaspi05} relation are shown for
reference.}

\label{fig:rLrelation}
\end{figure}

\clearpage

\begin{figure}
\epsscale{1}
\plotone{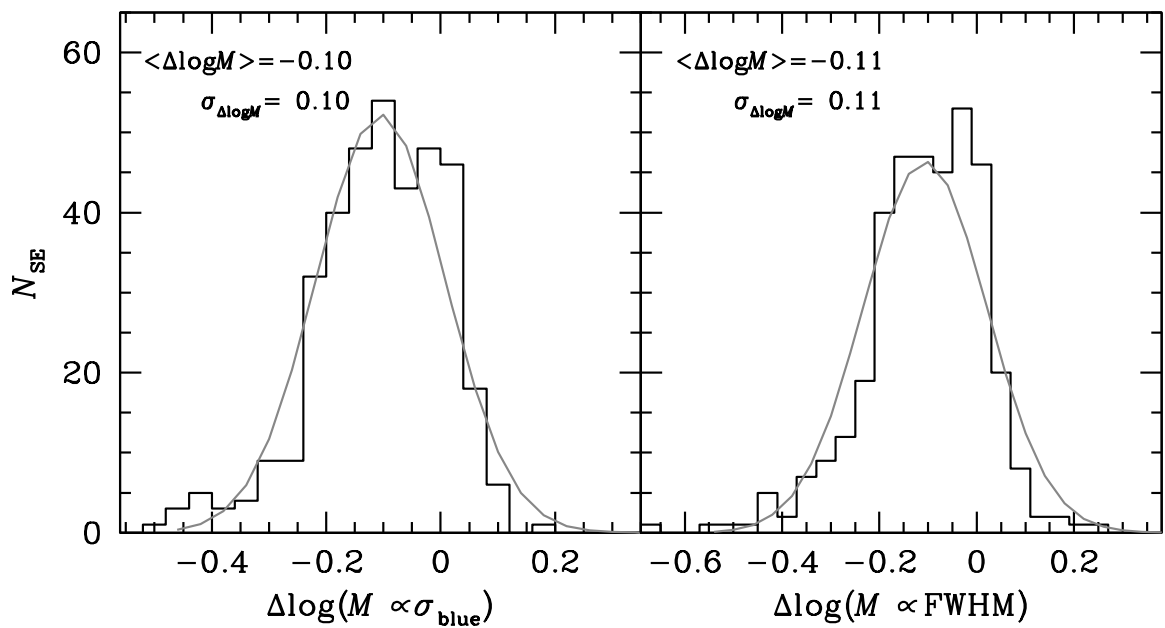}

\caption{Distributions of the differences between each SE mass in a given
observing year and the reverberation virial mass from that same year,
plotted for masses calculated with \sigbl\ (left) and FWHM (right).
Mass differences are shown for every spectrum in the full sample of
$370$ observations of NGC~5548 after subtraction of narrow emission-line
components and host starlight contribution.}

\label{fig:deltalogM}
\end{figure}
\clearpage

\begin{figure}
\epsscale{1}
\plotone{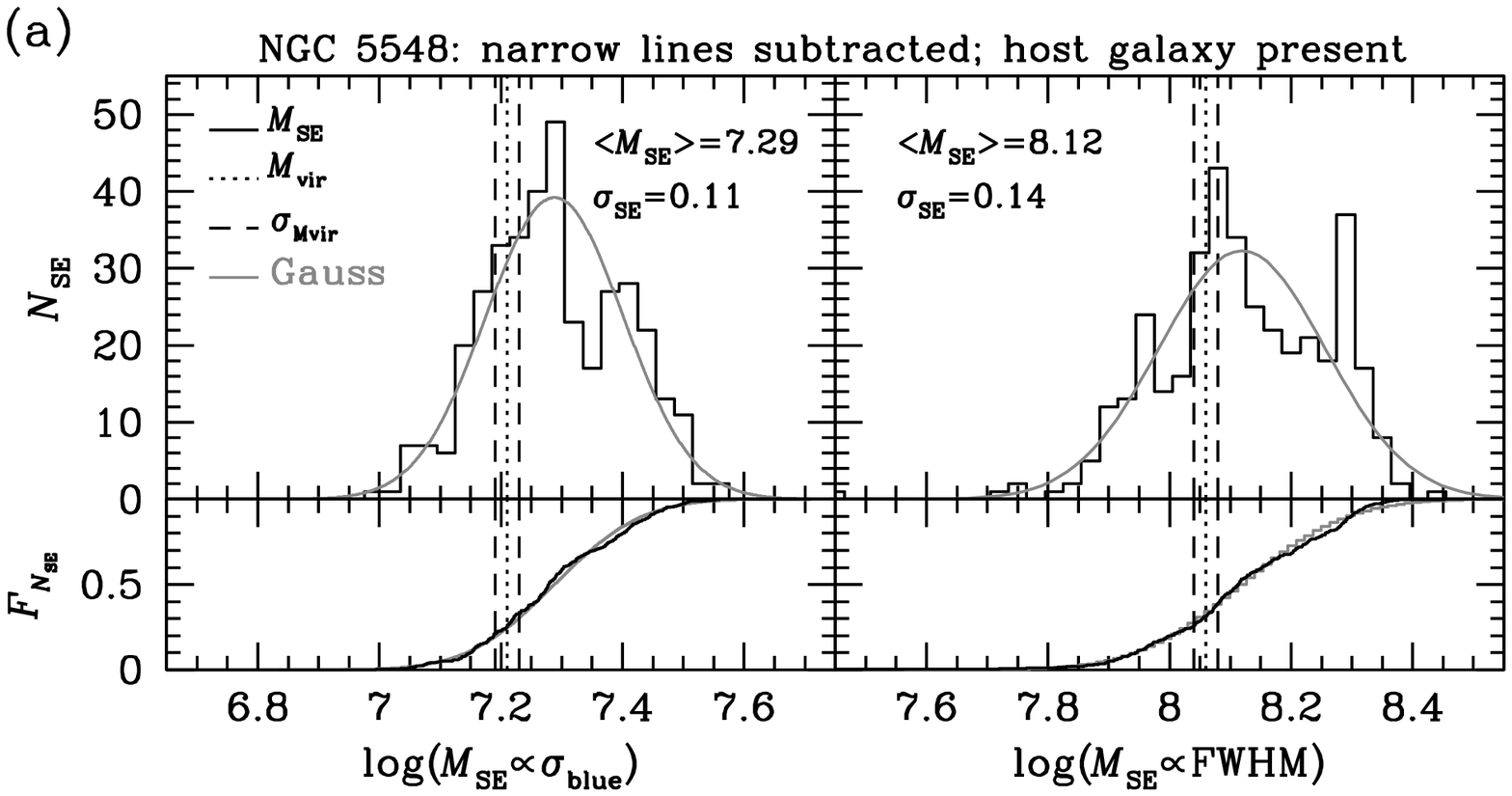}
\plotone{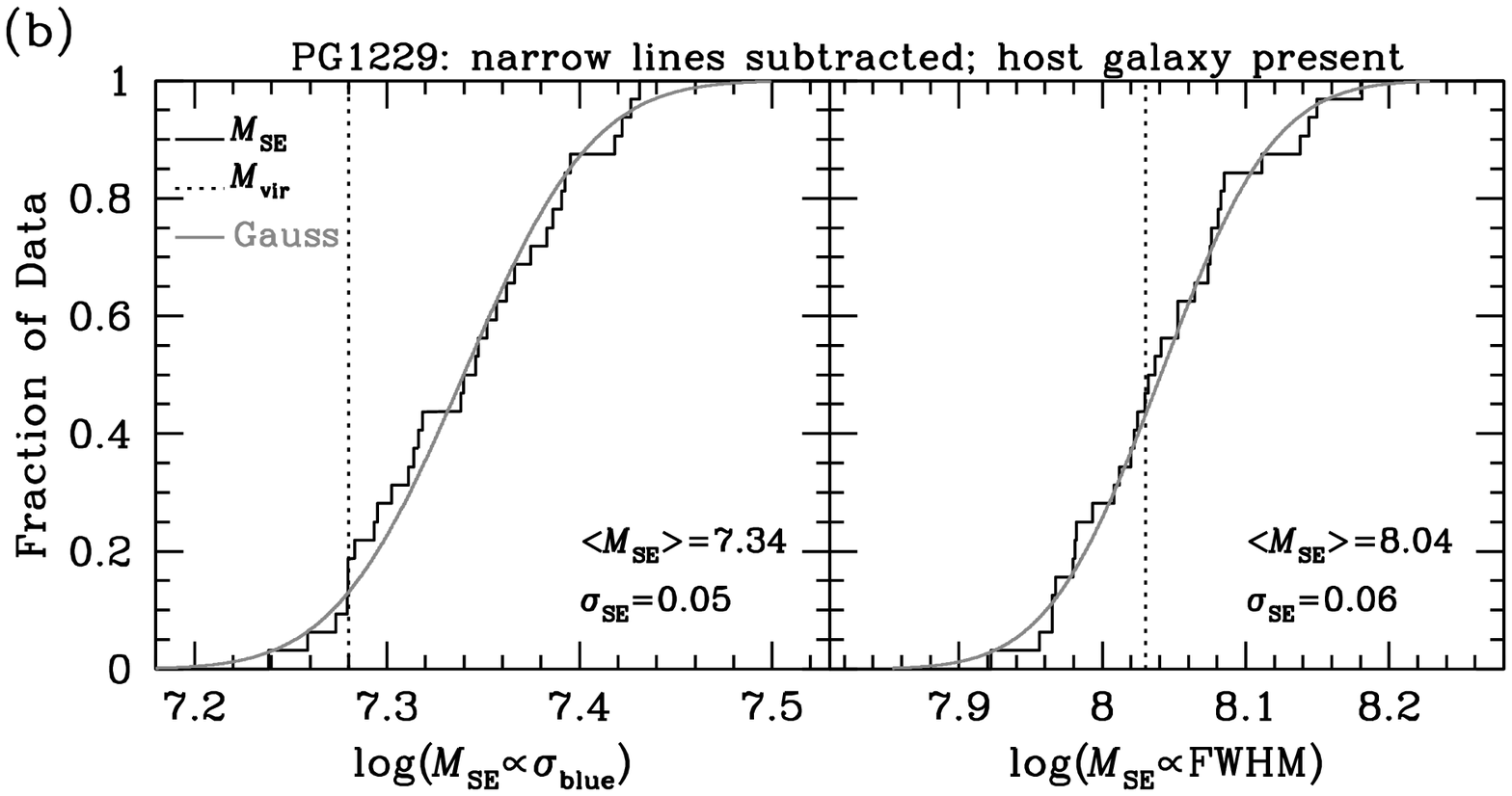}

\caption{Same as Figure \ref{fig:SEVPdistr_var}, except the host-galaxy
flux contribution has not been removed.  The narrow-line components have
been subtracted from the spectra before measuring the \Hbeta\ line width
and calculating the black hole mass.}

\label{fig:SEVPnl_nogs}
\end{figure}

\begin{figure}
\epsscale{1}
\plotone{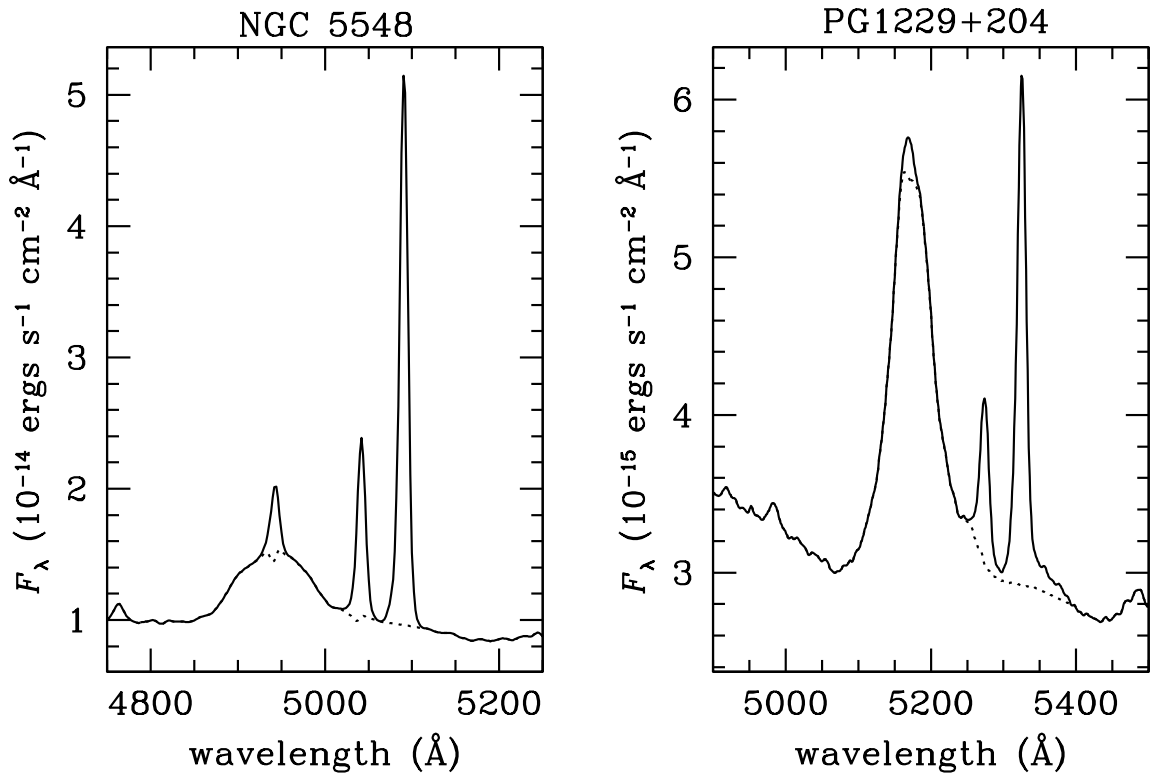}

\caption{Mean spectra of NGC~5548 (left) and PG1229 (right) with narrow
emission lines (solid) and after subtraction of the narrow emission
lines (dotted).}

\label{fig:meanspec}
\end{figure}

\begin{figure}
\epsscale{1}
\plotone{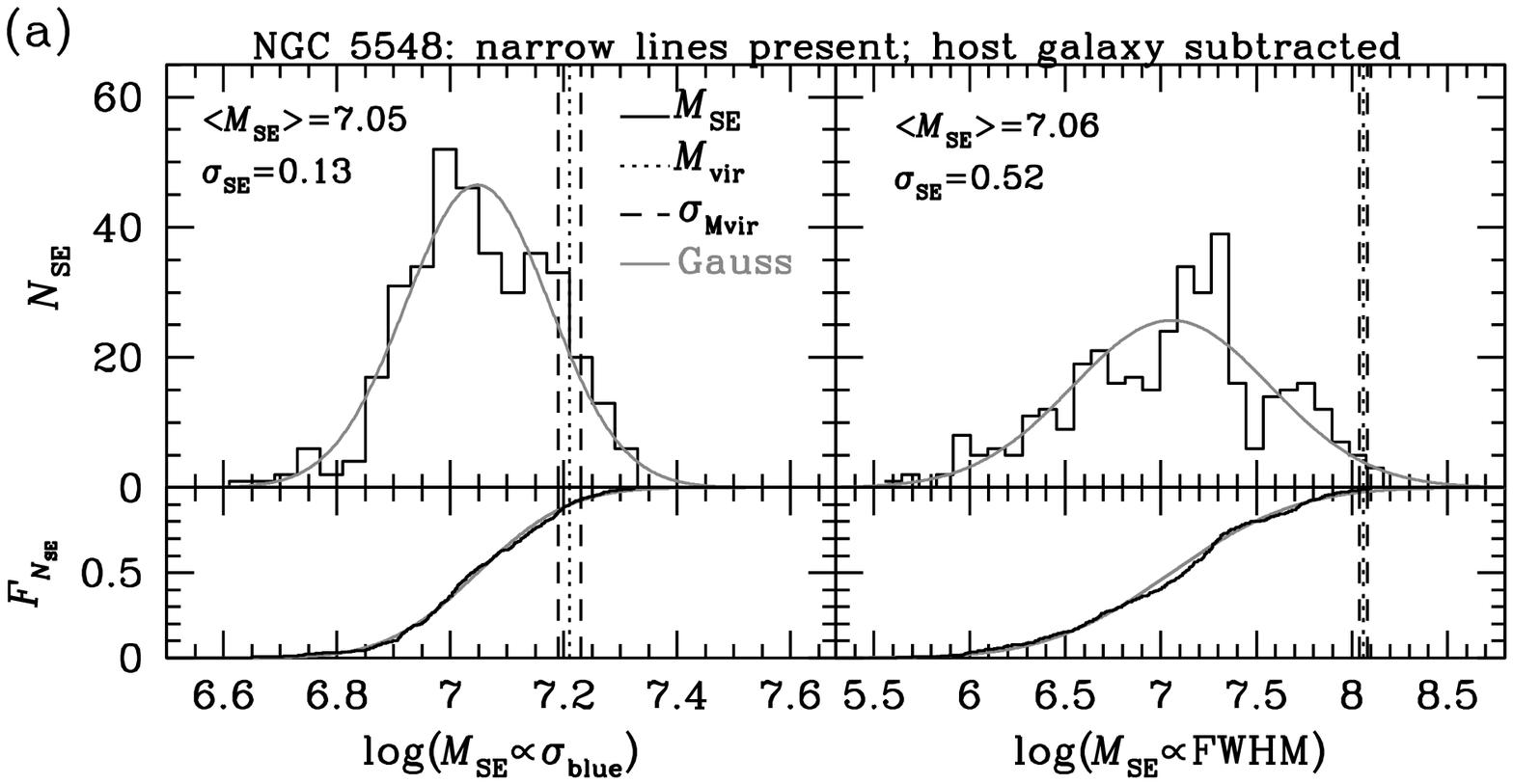}
\plotone{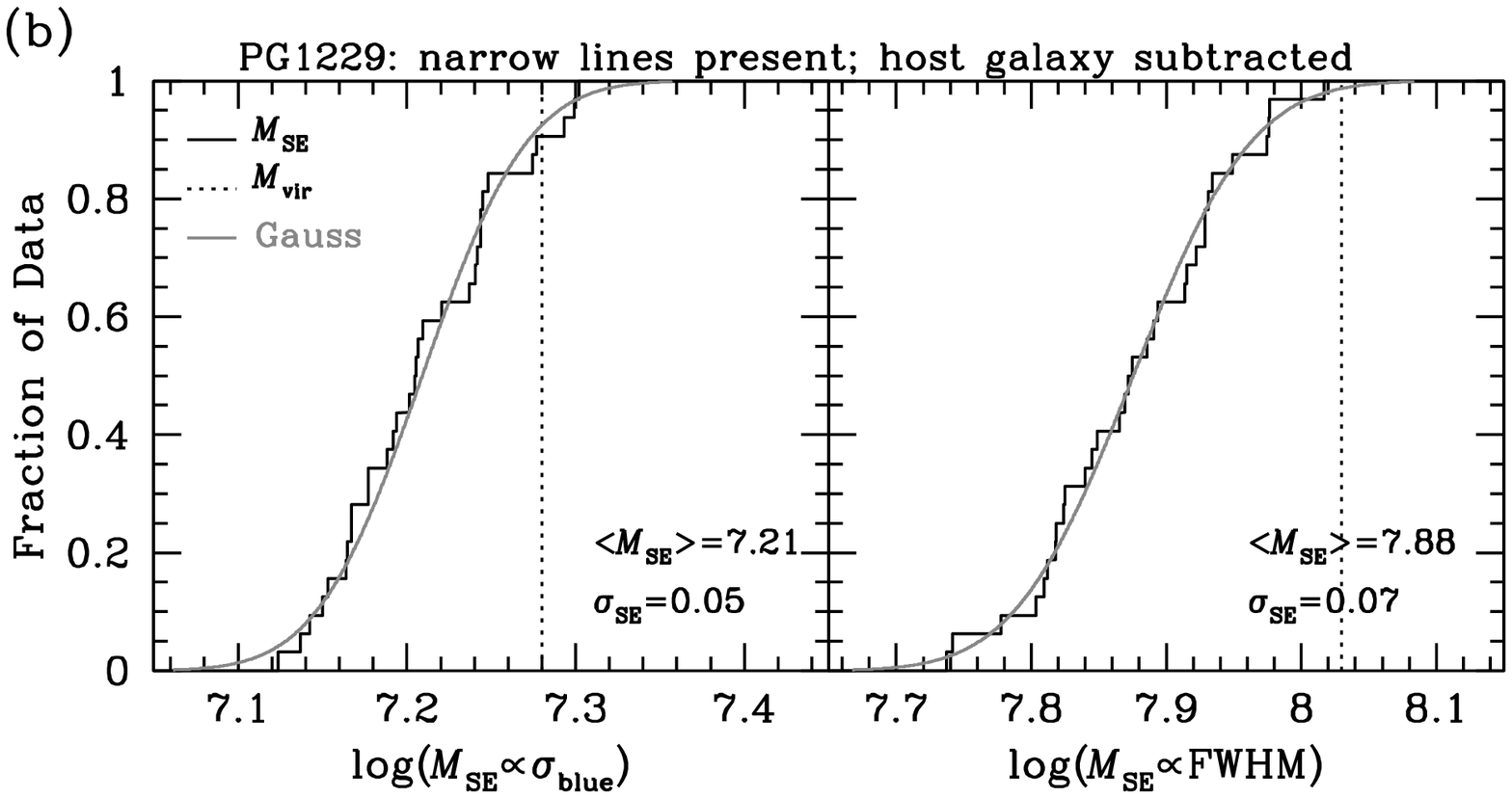}

\caption{Same as Figure \ref{fig:SEVPdistr_var}, except the narrow
emission lines have not been removed from the spectra before measuring
the \Hbeta\ line width.  The host-galaxy contribution to the flux was
removed before determination of the masses.}

\label{fig:SEVPdistr_wlgs}
\end{figure}

\begin{figure}
\epsscale{1}
\plotone{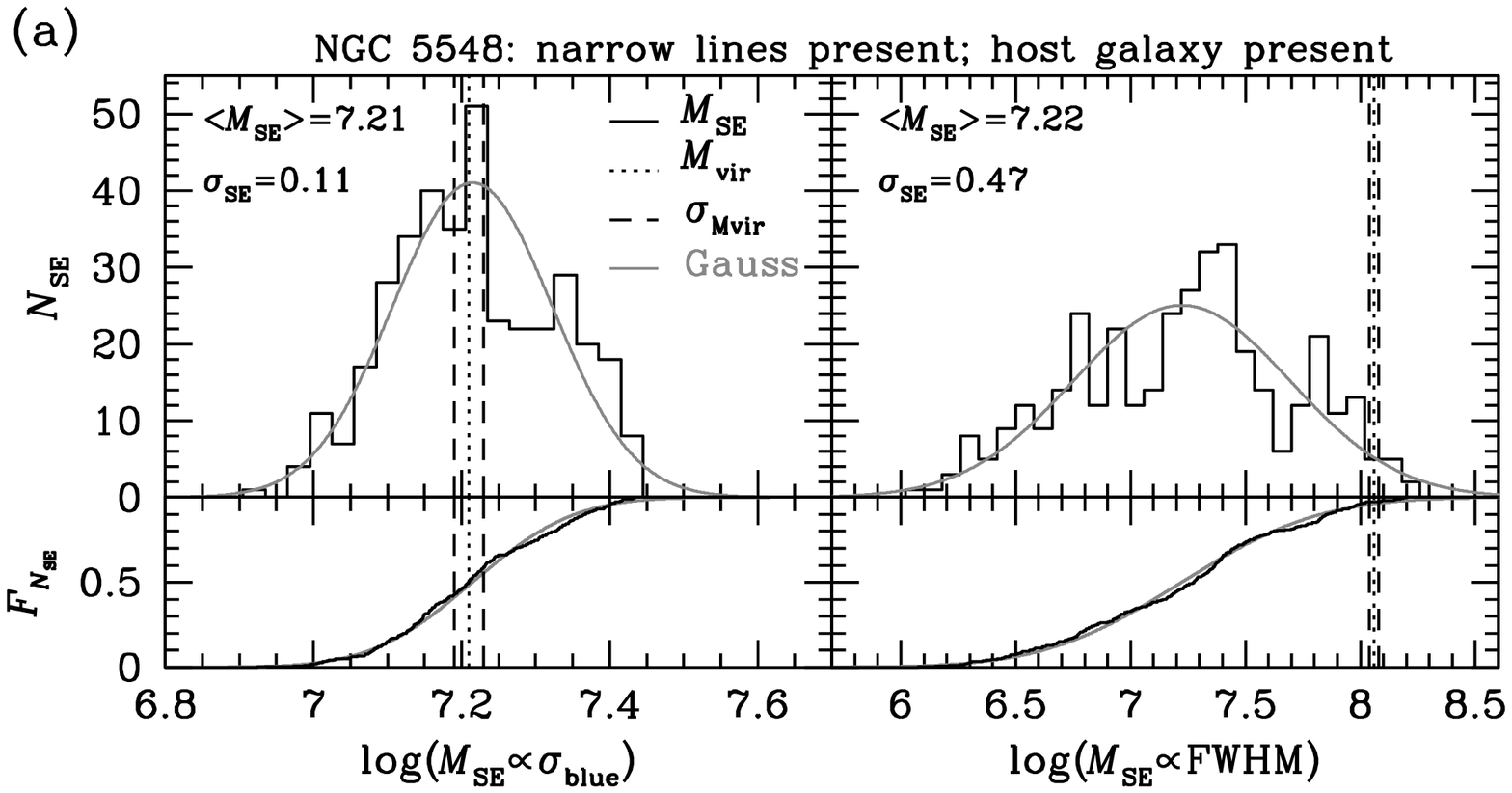}
\plotone{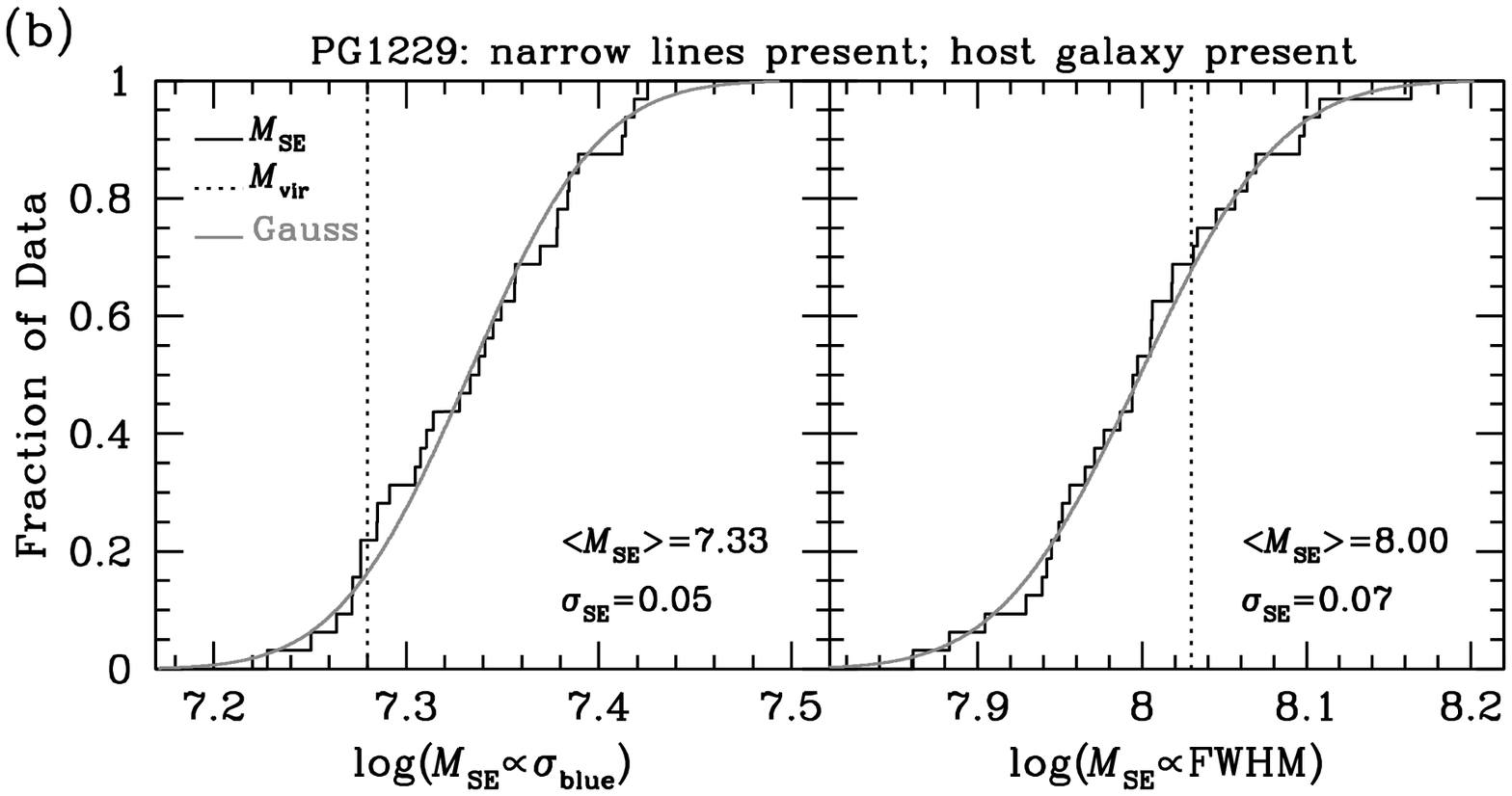}

\caption{Same as Figure \ref{fig:SEVPdistr_var}, except neither the
narrow-line components nor the host-galaxy flux contribution have been
removed before determination of the masses.}

\label{fig:SEVP_nocorrections}
\end{figure}

\begin{figure}
\epsscale{1}
\plotone{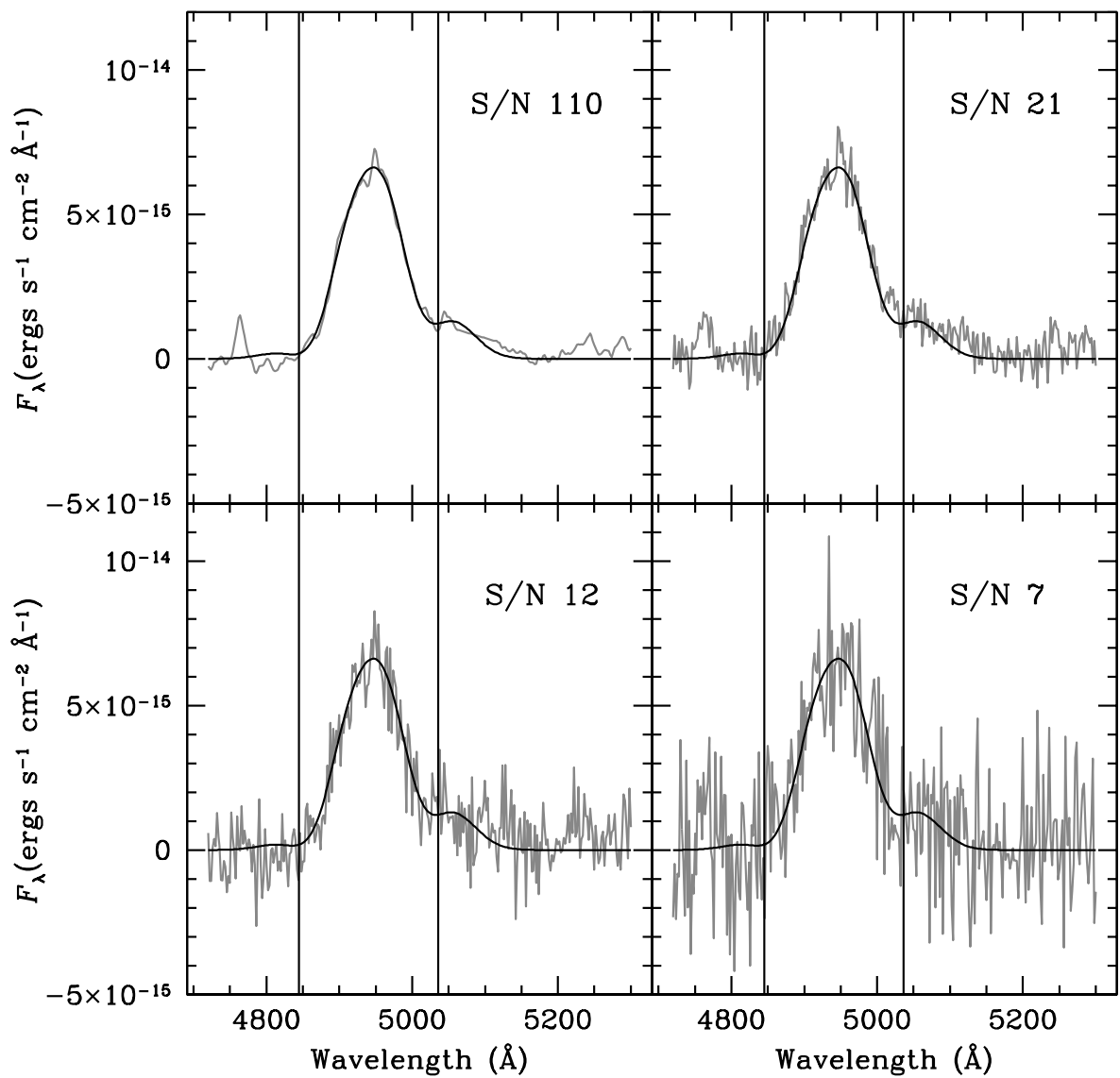}

\caption{Example spectrum of NGC~5548 (top left) and three artificial
degradations of the same spectrum with the resultant $S/N$ labeled for
each. The solid black lines are the Gauss-Hermite polynomial fits to
each spectrum (gray lines) as described in \S \ref{S_VPs_from_fitsSN}.
The vertical lines show the assumed boundaries of the broad H$\beta$
line used for measuring the line widths from both the actual data and
the Gauss-Hermite fits.}

\label{fig:SNdegrade}
\end{figure}

\begin{figure}
\epsscale{1}
\plottwo{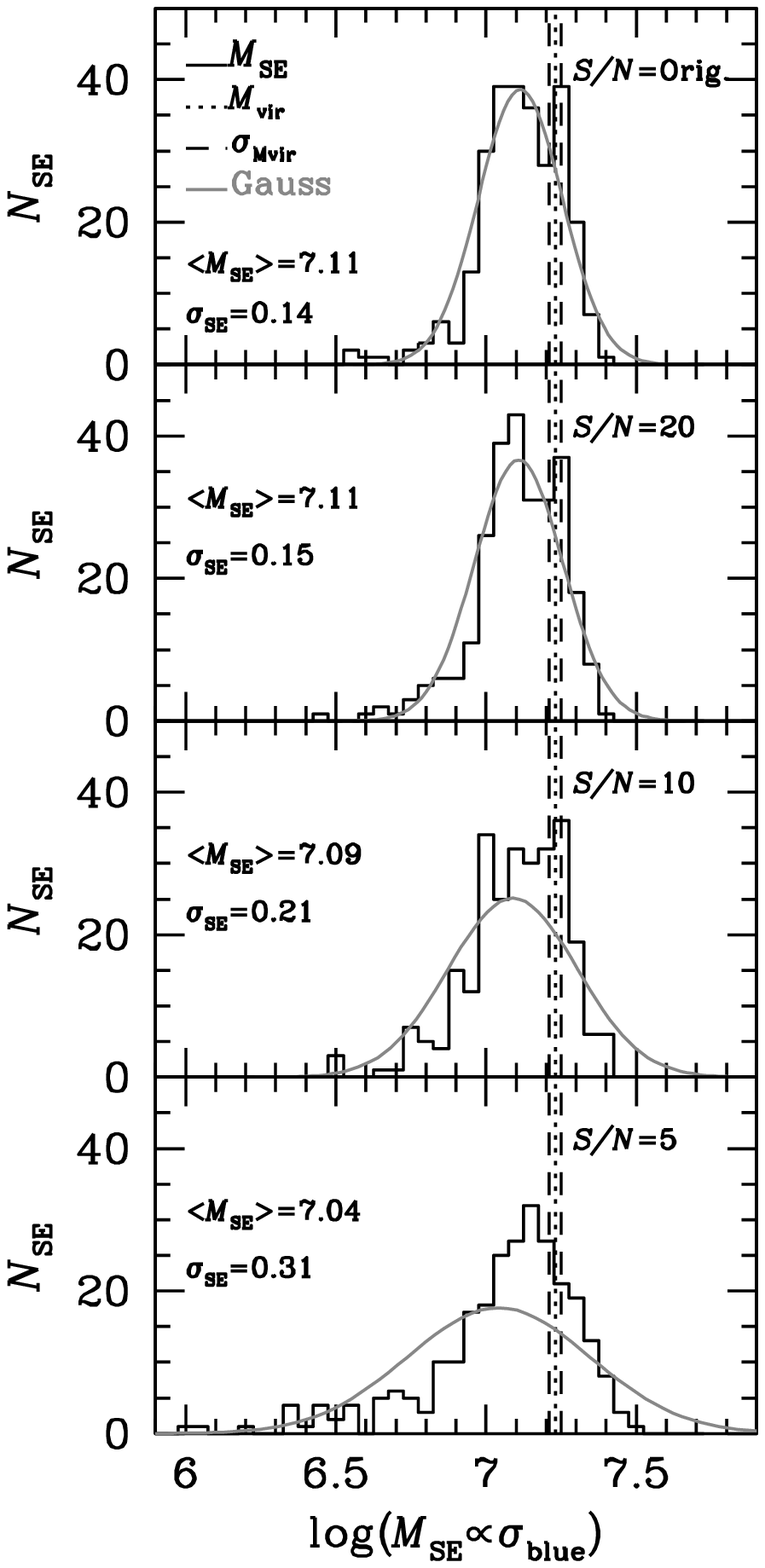}{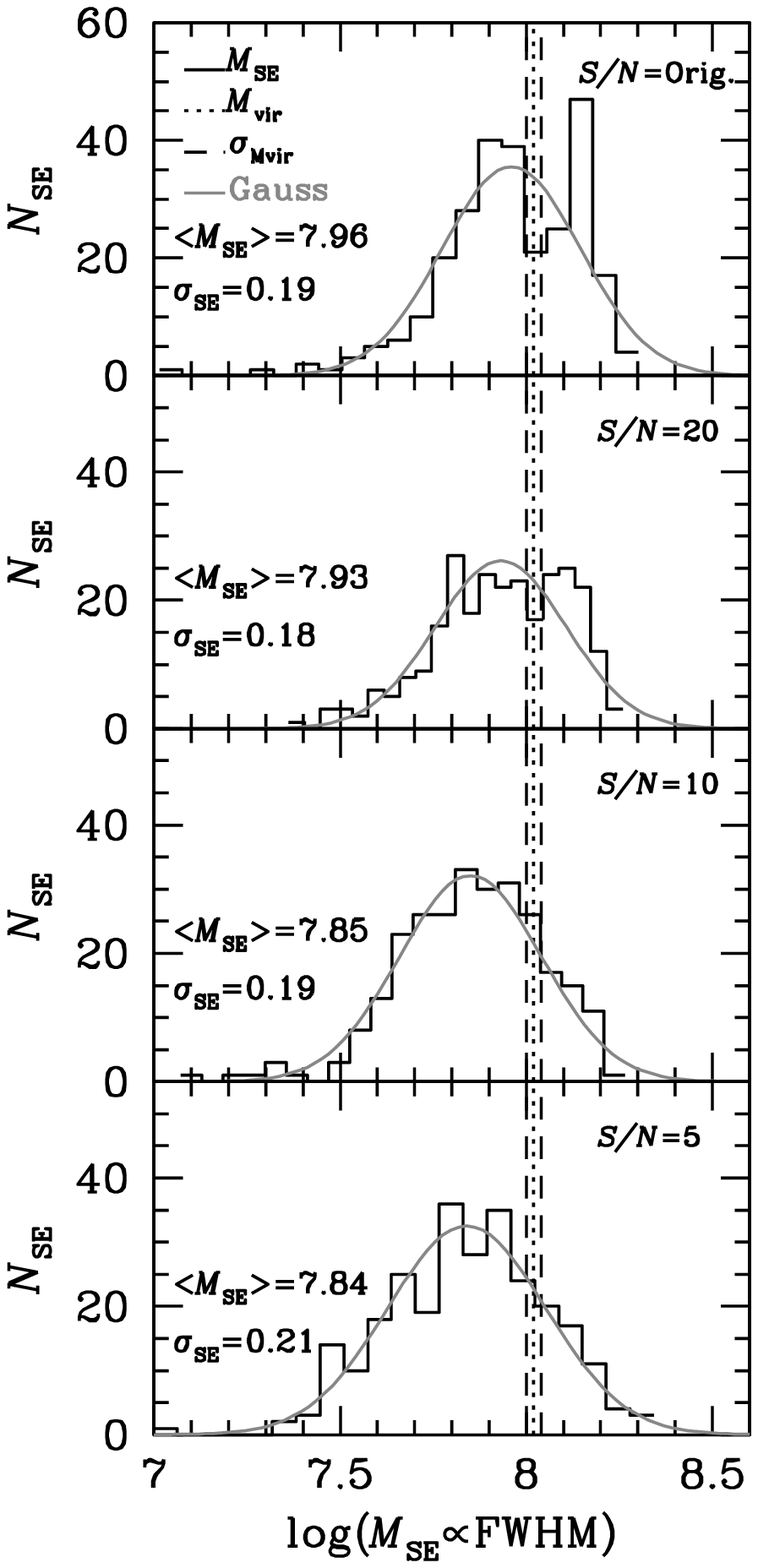}

\caption{Virial mass distributions for the original and $S/N$ degraded
NGC~5548 Perkins data set, using line widths measured directly from the
data.  Virial masses are calculated in all cases by measuring the
velocity dispersion of the broad H$\beta$ emission from narrow
line-subtracted spectra using \sigbl\ (left) and FWHM (right).  All
virial masses are calculated using a value of $L_{5100}$ that has been
corrected for host galaxy starlight.}

\label{fig:SEVPdataSN}
\end{figure}
\clearpage

\begin{figure}
\epsscale{1}
\plottwo{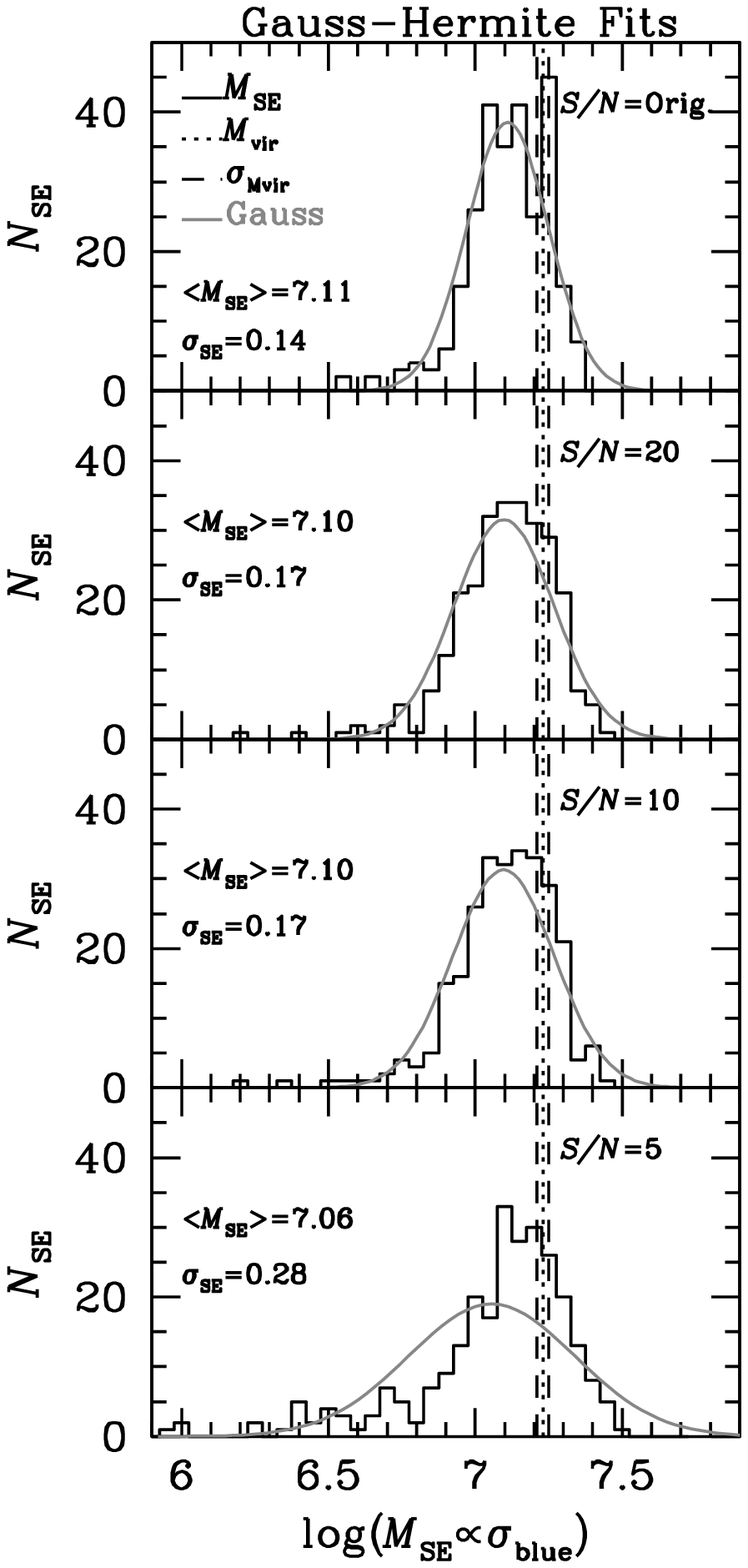}{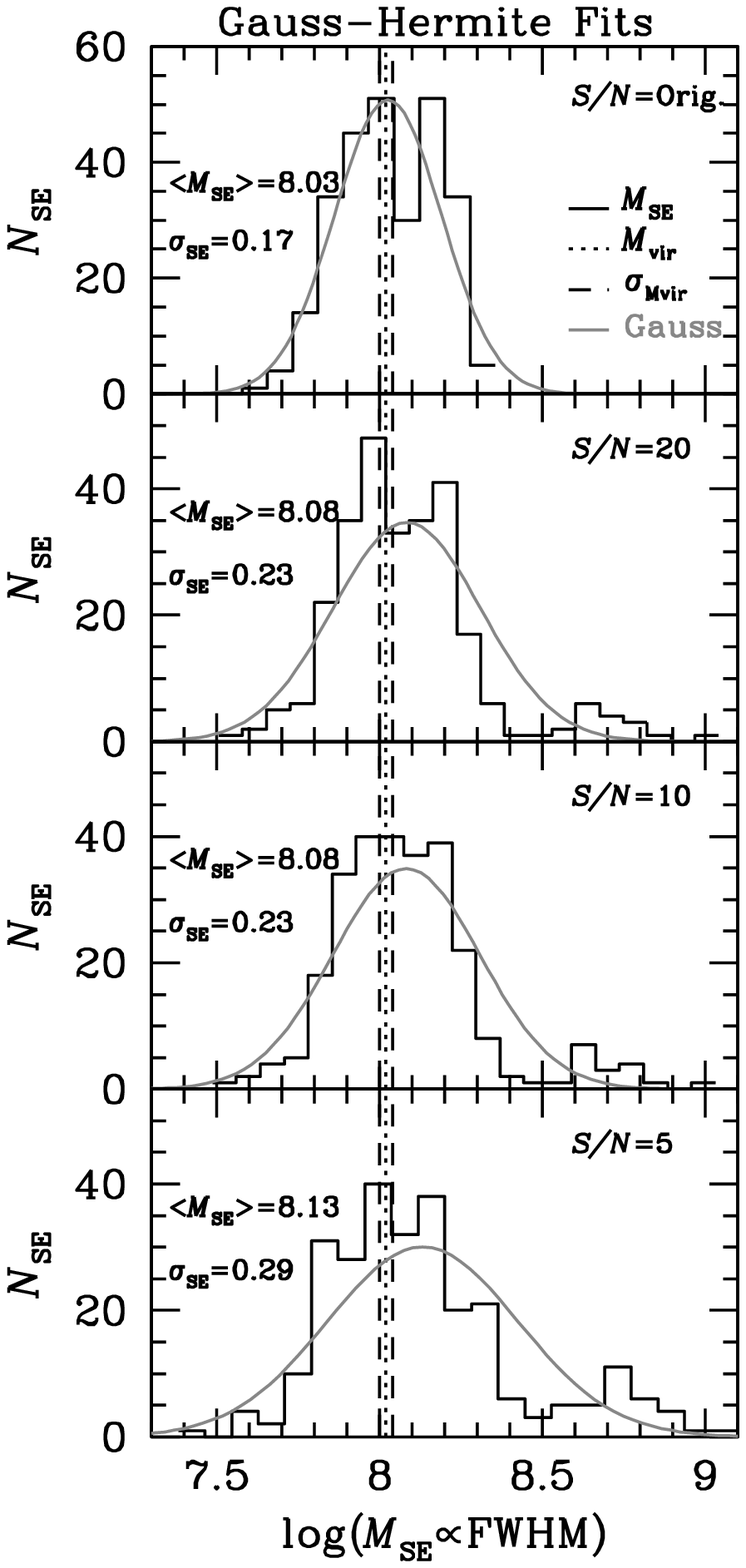}

\caption{Virial mass distributions based on sixth order Gauss-Hermite
polynomial fits to the original and $S/N$ degraded NGC~5548 Perkins data
set.  Virial masses are calculated by measuring the velocity dispersion
of the Gauss-Hermite polynomial fit to narrow line-subtracted broad
H$\beta$ emission line with \sigbl\ (left) and the FWHM (right).  All
virial masses are calculated with a value of $L_{5100}$ that has been
corrected for host galaxy starlight.}

\label{fig:SEVPfitsSN}
\end{figure}
\clearpage

\begin{figure}
\epsscale{1}
\plotone{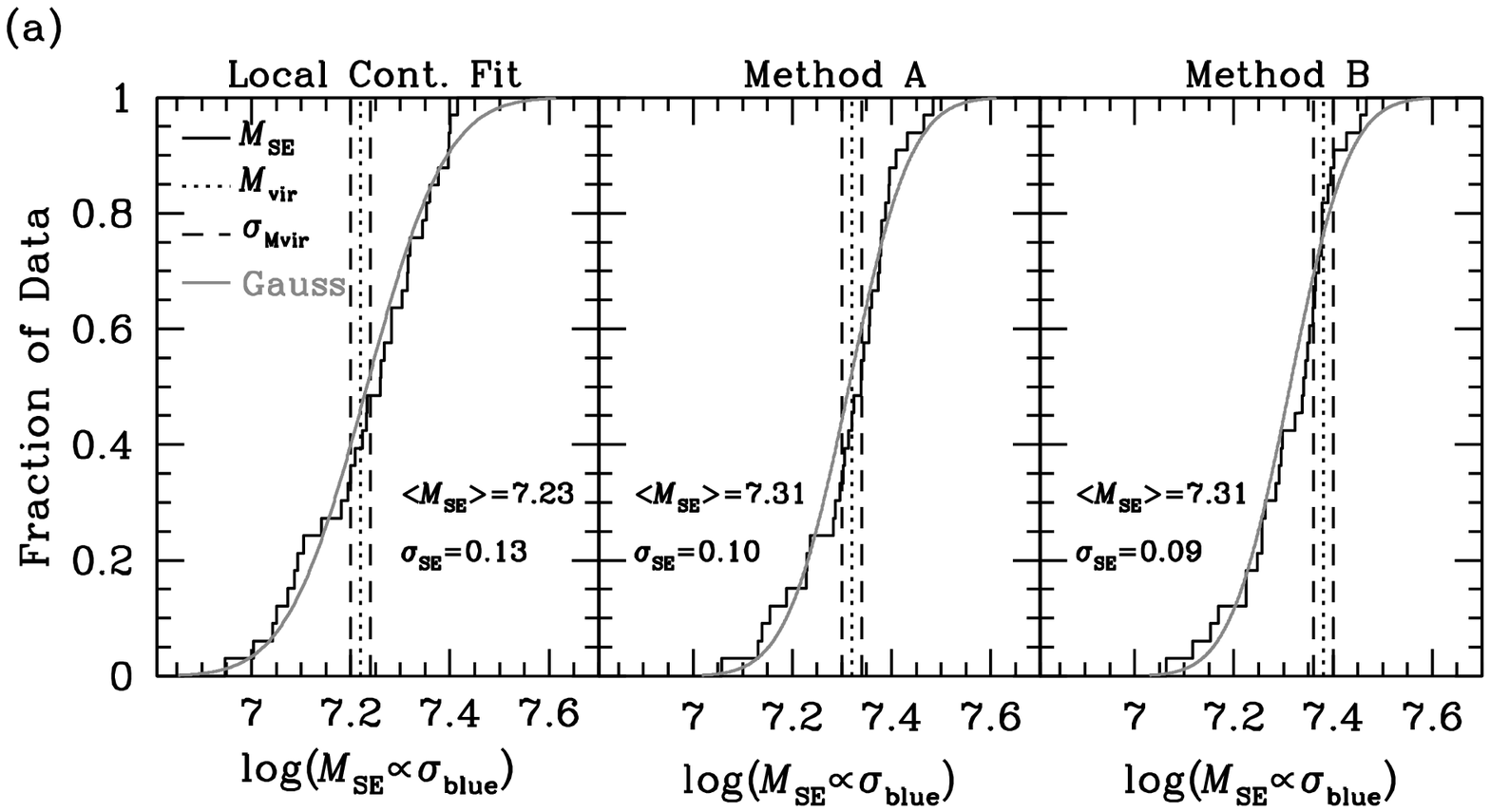}
\plotone{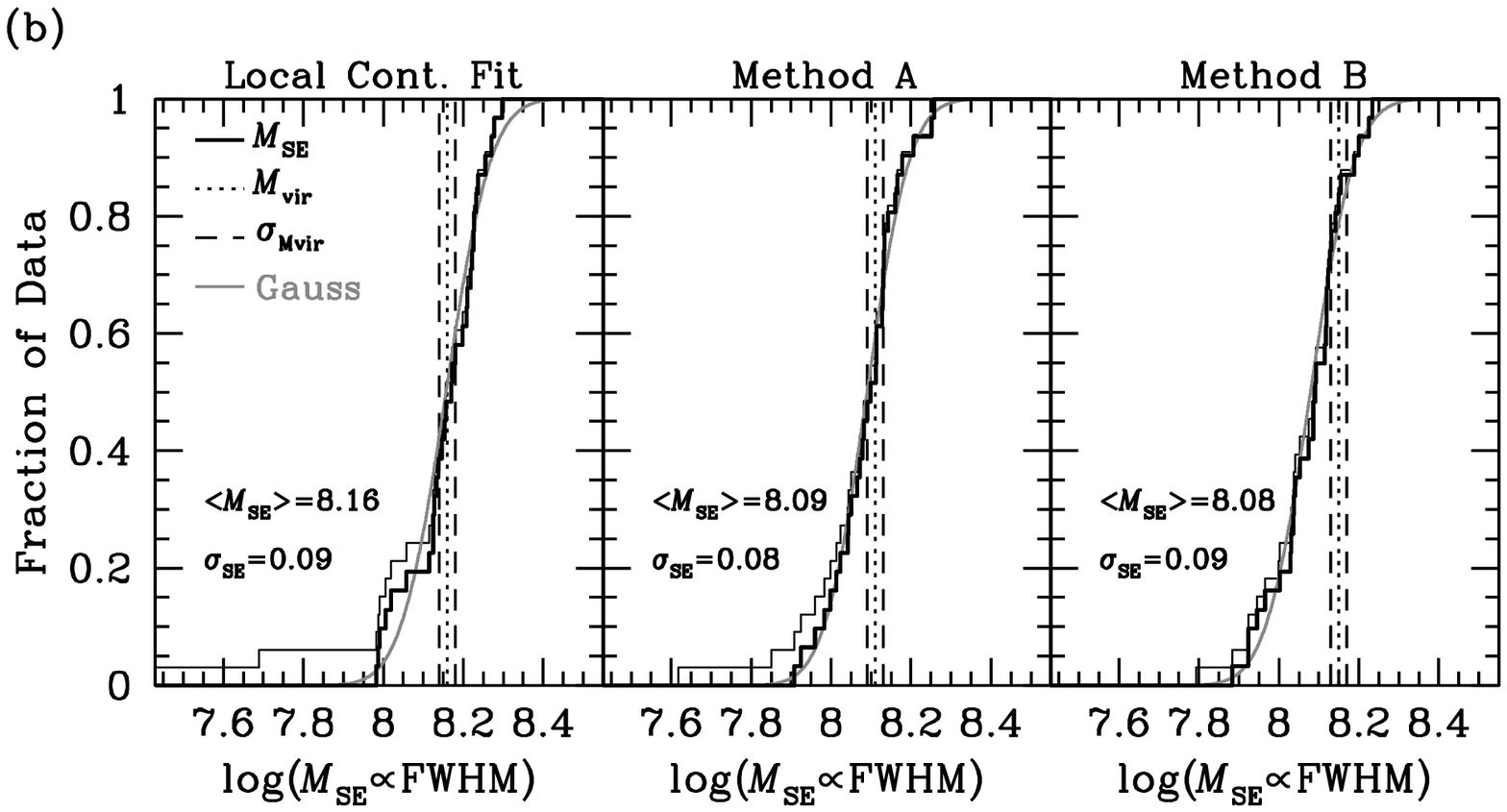}

\caption{Cumulative distribution functions of NGC~5548 SE virial masses
for decomposition data sets with \sigbl\ (a) and FWHM (b).  The panels
show virial mass distributions that have been calculated using a local
continuum fit to the continuum underneath the \Hbeta\ line (left),
spectral decomposition method A (middle), and spectral decomposition
method B (right).  Statistics listed in the bottom three panels do not
include the outliers plotted in the thin black line, as described in
\S\ref{S_Res_decomp}.}

\label{fig:SEVPdistr_decomp}
\end{figure}
\clearpage

\begin{figure}
\epsscale{1}
\plotone{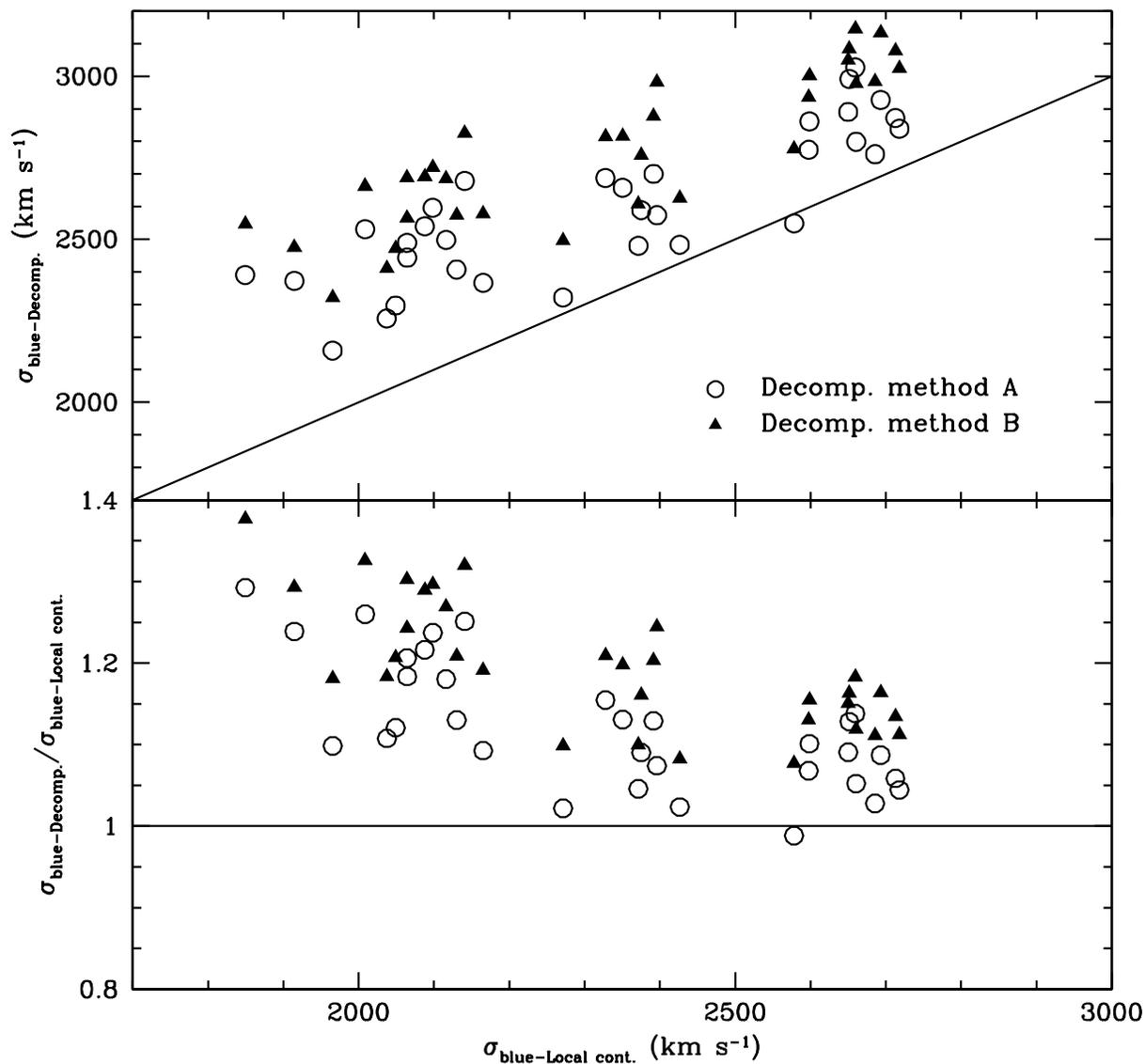}

\caption{Comparison of the \Hbeta\ line dispersion measurements (\sigbl)
using the various techniques described in \S\ref{S_Res_decomp} to
account for the continuum flux level under the emission line.  The open
circles represent the line width measurements using decomposition method
A results, and the black triangles show results using method B.  In the
top panel, these values are plotted against widths measured using a
local continuum fit, and the bottom panel shows the residuals with
respect to the width from the local continua method.  The solid black
line in each panel shows a 1:1 correlation between the measured
\sigbl\ values.}

\label{fig:sigbl_plot}
\end{figure}
\clearpage

\begin{figure}
\epsscale{1}
\plotone{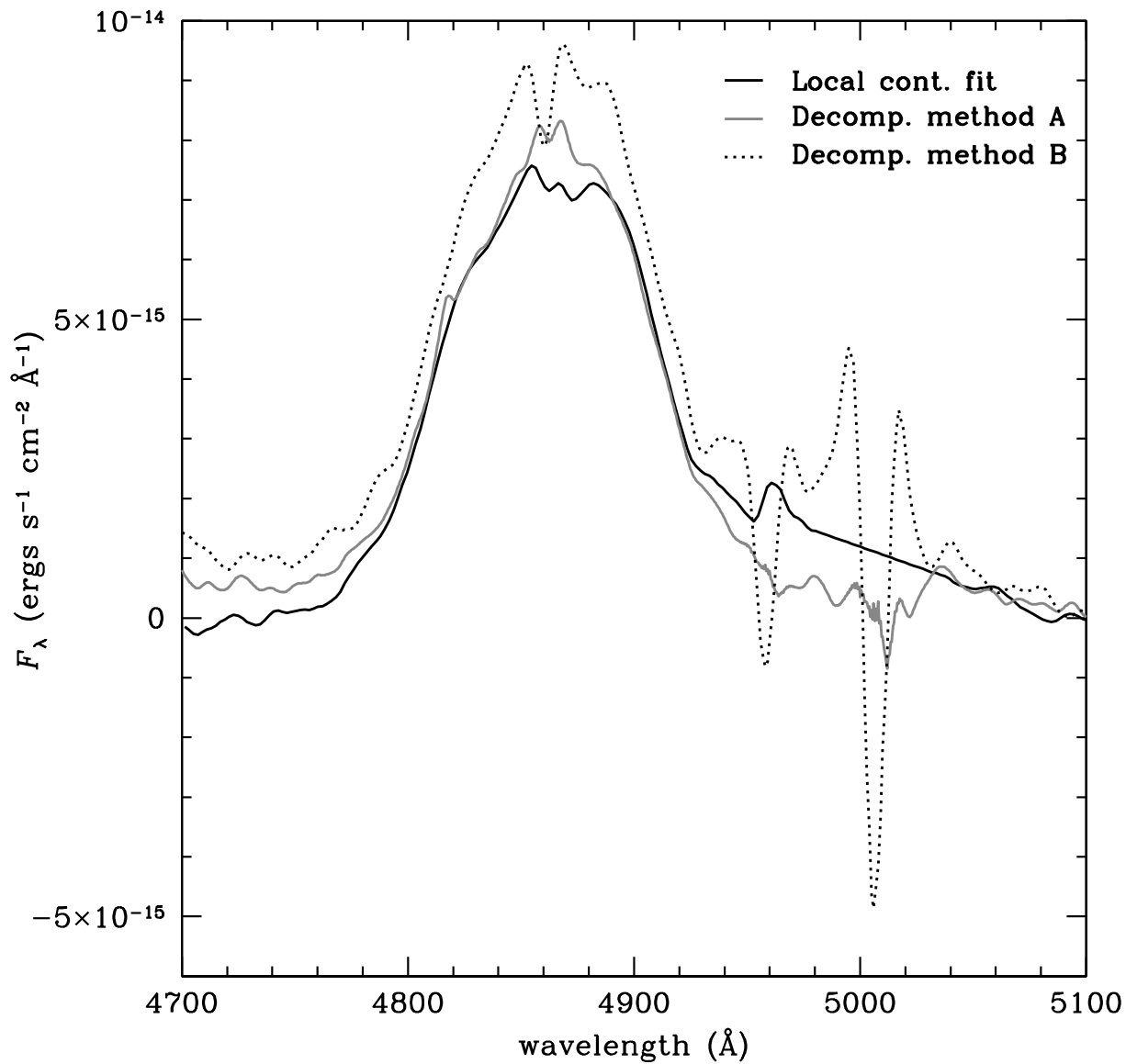}

\caption{Mean spectra of NGC~5548 created using three different
techniques to account for the continuum flux level under the \Hbeta\
emission line (as described in \S \ref{S_Res_decomp}).  The black line
shows the mean spectrum formed after simply fitting and subtracting a
local linear continuum to each of the 33 spectra.  The gray line shows
the mean continuum-subtracted spectrum formed after deblending the
spectral components from the same 33 spectra with method A, and the
dotted line is the same, but for method B.}

\label{fig:profile_oplot}
\end{figure}
\clearpage

\begin{figure}
\epsscale{1}
\plotone{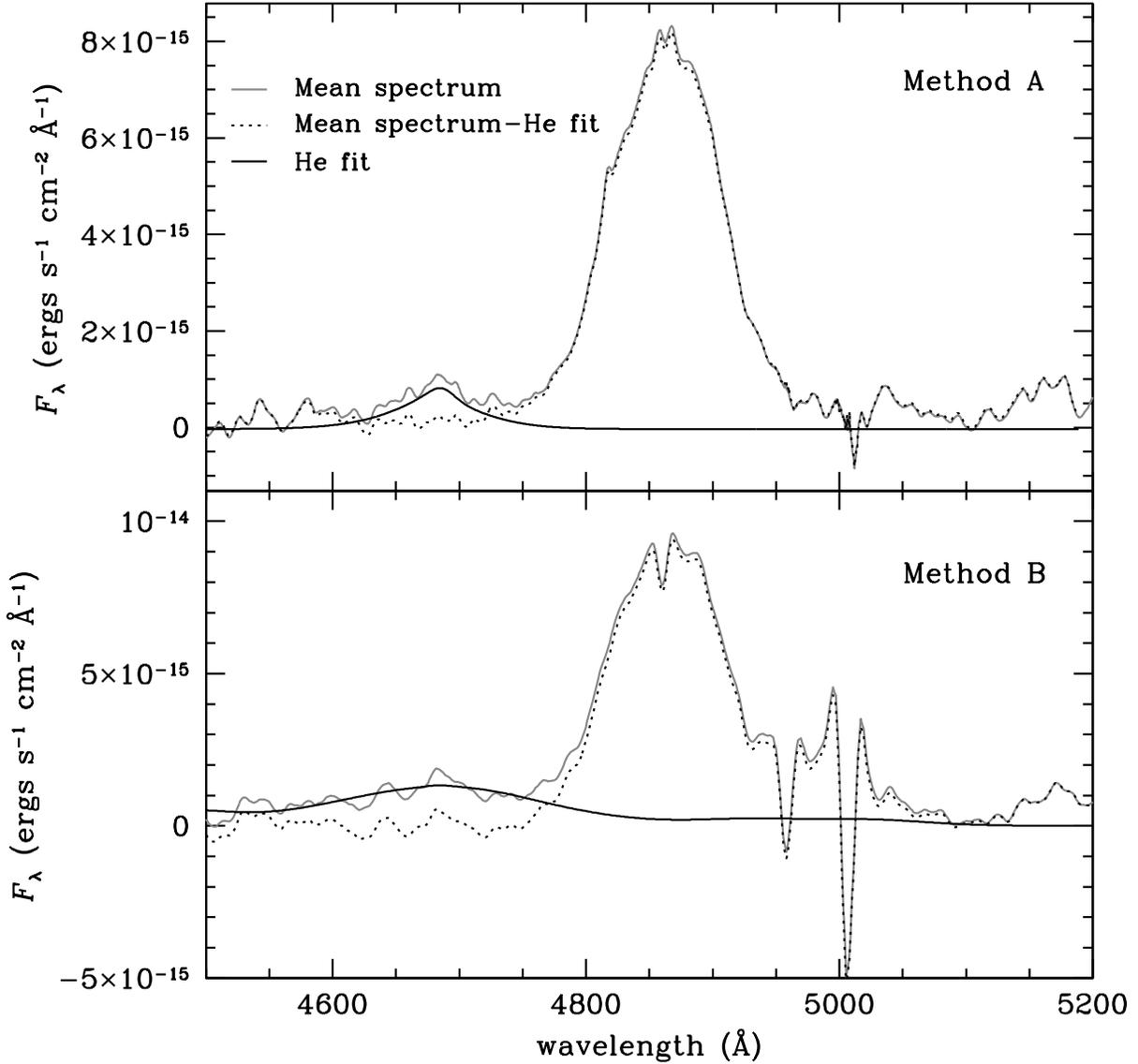}

\caption{Mean spectra of NGC~5548 both before (gray lines) and after
(dotted lines) subtracting the \HeII\ emission line (black lines). The
top panel show the results from decomposition method A, where the \heii\
line was modeled assuming the line profile of \Hbeta\ is the same as
\Halpha.  The bottom panel shows the results from decomposition method
B, where the \heii\ line is fit assuming the same velocity width as the
unblended \HeI\ line.}

\label{fig:Hefits}
\end{figure}
\clearpage

\begin{figure}
\epsscale{1}
\plotone{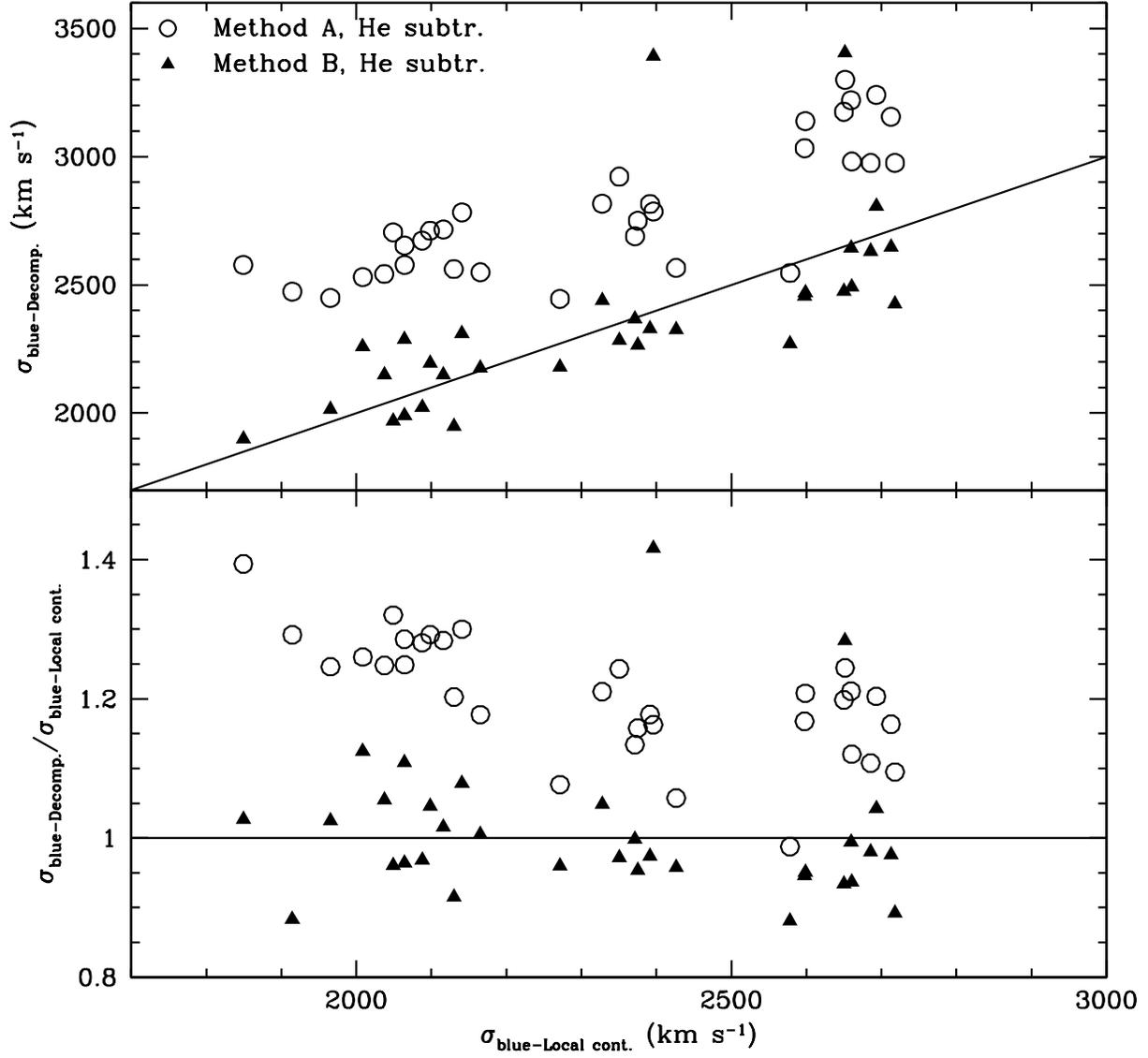}

\caption{Same as Figure \ref{fig:sigbl_plot}, except the \HeII\ emission
line has been subtracted from each of the spectra before measuring the
\Hbeta\ line dispersion (\sigbl).}

\label{fig:noHe_sigbl_plot}
\end{figure}
\clearpage

\begin{figure}
\epsscale{1}
\plotone{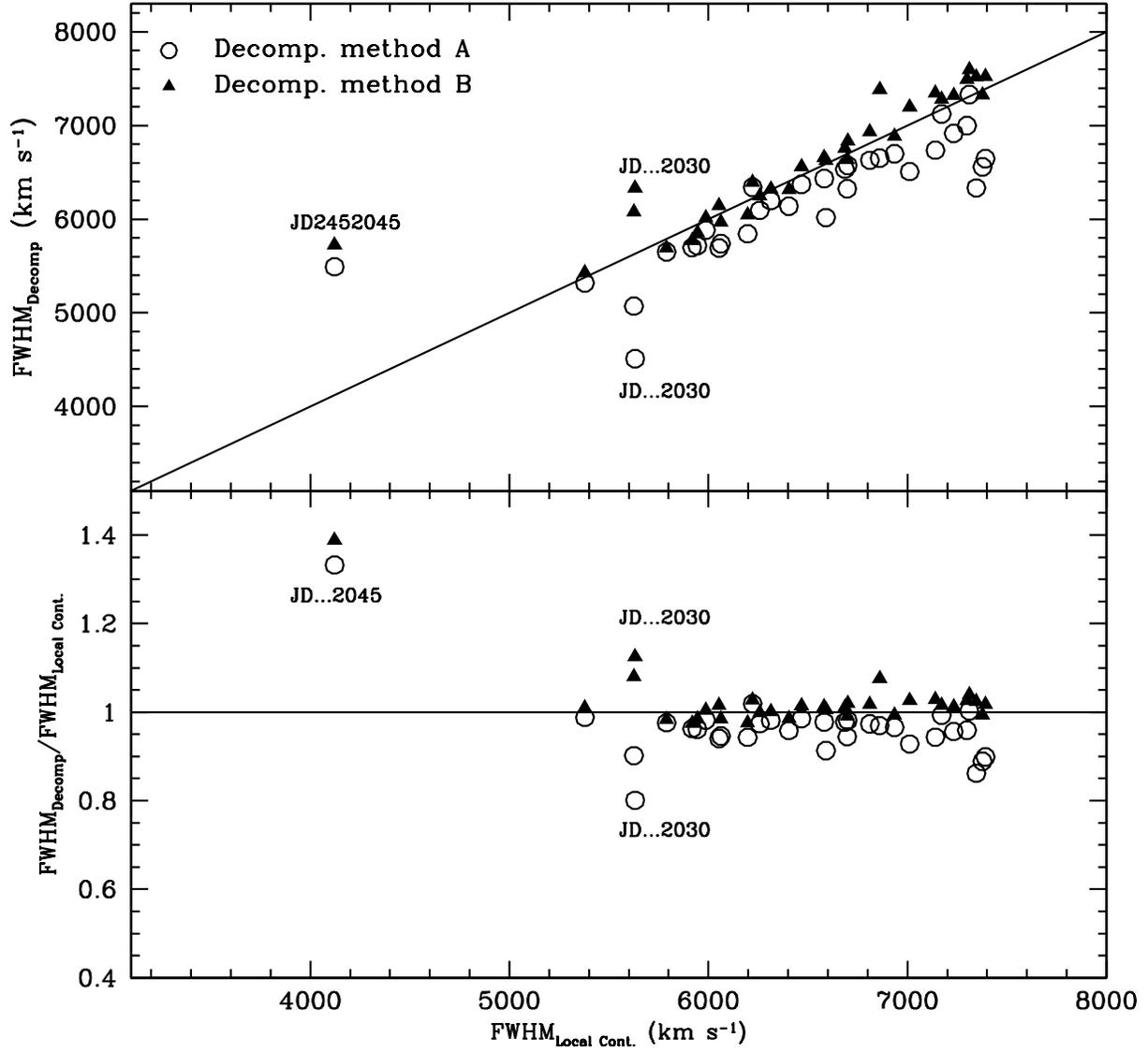}

\caption{Same as Figure \ref{fig:sigbl_plot}, except the width of the
\Hbeta\ line is characterized here by FWHM rather than \sigbl.  Outliers
discussed in \S \ref{S_Res_decomp} are individually labeled by Julian
Date.}

\label{fig:fwhm_plot}
\end{figure}
\clearpage

\begin{figure}
\epsscale{1}
\plotone{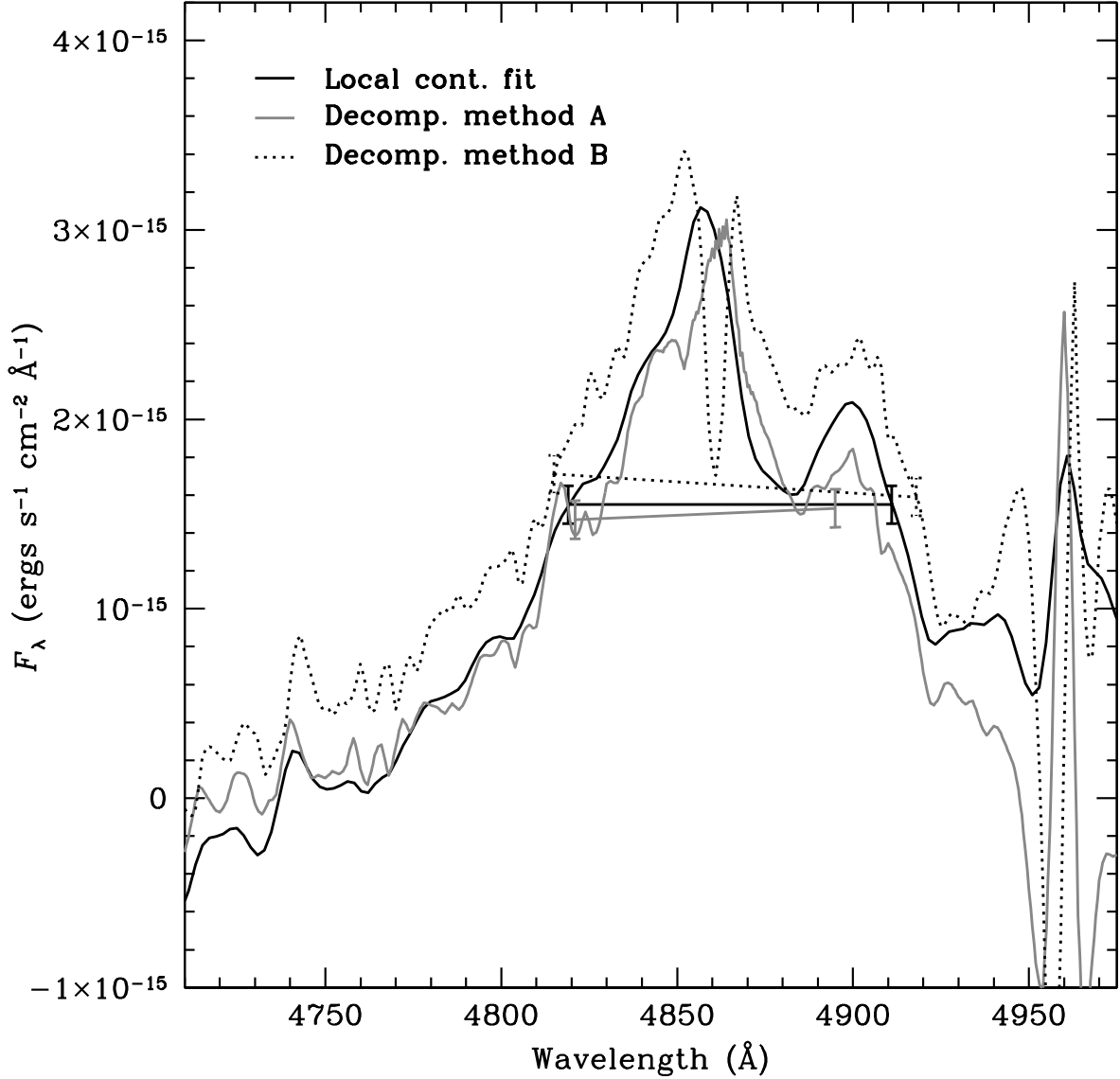}

\caption{FWHM measurements of the \Hbeta\ broad line for JD2452030 using
each of the three spectral analysis methods discussed in
\S\ref{S_Res_decomp}.  The spectra shown here illustrate complications
that arise when using FWHM to characterize a complex emission line
structure, especially when the narrow emission-line components are not
well-determined.  The \Hbeta\ profile after subtracting a local
continuum fit (black line) has FWHM$=5632$\,\kms, the profile determined
from decomposition method A (gray line) has FWHM$=4511$\,\kms, and the
profile from decomposition method B (dotted line) has
FWHM$=6334$\,\kms.}

\label{fig:outlier_oplot}
\end{figure}
\clearpage

\end{document}